\documentclass[aps,prl,twocolumn]{revtex4-2}
\usepackage{epsfig}
\usepackage{amsfonts}
\usepackage{amssymb}
\usepackage{enumitem}
\usepackage{amsmath}
\usepackage{amsthm}
\usepackage{subfig}
\usepackage{extarrows}
\usepackage{bm}
\usepackage{hyperref}
\usepackage{mathrsfs}
\usepackage{bbm}
\usepackage{cancel}
\usepackage[dvipsnames]{xcolor}
\usepackage{accents}
\usepackage{relsize}
\usepackage[capitalize]{cleveref}
  
  \crefname{section}{Sec.}{Secs.}
  \crefname{appendix}{App.}{Apps.}

\usepackage{xcolor}
\usepackage{hyperref}
\usepackage{mathtools}
\usepackage{amssymb,amsmath,graphics}   
\newcommand{\eq}{\begin{equation}}
\newcommand{\eqe}{\end{equation}}
\newcommand{\eqa}{\begin{eqnarray}}
\newcommand{\eqae}{\end{eqnarray}}

\newcommand{\sect}[1]{\noindent\textbf{#1.\textemdash}\ }

\begin{document}

\title{New Recursions for the Canonical Scalar-Scaffolded Yang-Mills Amplitude}
\author{Jeffrey V. Backus}
\email{jvabackus@princeton.edu}

\affiliation{Joseph Henry Laboratories, Princeton University, Princeton, NJ 08540, USA}

\begin{abstract}

The recently-developed ``scalar-scaffolding'' formulation of gluon amplitudes casts the Yang-Mills (YM) amplitude as a well-defined Laurent series expansion in scalar variables, valid for any spacetime dimension and helicity configuration. In this letter, we exploit this new perspective to develop conceptually novel methods of computing YM tree amplitudes. First, using standard gluon factorization to determine all terms with poles, we show how gauge invariance uniquely fixes the piece with no poles (the ``contact term'') from only terms that have a single pole. This allows us to write a YM recursion not only for the full amplitude but also for the amplitude up to any order in the Laurent series. Next, by imposing gauge invariance for terms with poles, we write down relations which compute numerators recursively in the amplitude's Laurent series expansion. Starting from an initial set of cuts depending only on the $(n-1)$-point amplitude, these formulae allow us to determine the remaining terms in the $n$-point amplitude. Finally, we use this ``Laurent series recursion'' to derive a recursion solely for the contact term. We speculate on the possibility that this and analogous recursions for any term in the amplitude may be solved. In attached Mathematica notebooks, we give implementations of these three recursions.

\end{abstract}
\maketitle

\sect{Introduction} Recent years have seen the emergence of a new perspective on the scattering of colored, non-supersymmetric particles \cite{Arkani-Hamed:2023lbd,Arkani-Hamed:2023mvg,Arkani-Hamed:2023swr,Arkani-Hamed:2023jry,Arkani-Hamed:2024nhp,Arkani-Hamed:2024vna,Arkani-Hamed:2024yvu,Arkani-Hamed:2024fyd,Arkani-Hamed:2024tzl,Arkani-Hamed:2024pzc,Backus:2025njt,Cao:2025lzv,De:2024wsy,Salvatori:2025oib}. This formulation --- inspired originally by the discovery of positive geometry in Tr$(\phi^3)$ theory \cite{Arkani-Hamed:2017mur,Arkani-Hamed:2019vag} --- has shed light on a number of hidden properties of amplitudes \cite{Rodina:2024yfc,Backus:2025hpn,Jones:2025rbv,Paranjape:2025wjk,Feng:2025dci,Li:2025suo,Zhou:2024ddy}, spreading to uncolored theories \cite{Bartsch:2024amu,Zhou:2025tvq,Li:2024qfp} and even to the cosmological wavefunction \cite{De:2025bmf,Figueiredo:2025daa,Arkani-Hamed:2024jbp}. Among these many developments is the idea of ``scalar-scaffolding'' the Yang-Mills (YM) amplitude \cite{Arkani-Hamed:2023jry}, where we imagine that each external gluon was produced by two colored scalars. In doing so, we trade the polarization vectors $\epsilon_i^\mu$ and gluon momenta $q_i^\mu$ for scalar momenta $p_i^\mu$ and write the YM amplitude purely in terms of Lorentz-invariant scalar variables $X_{i,j} = (p_i + p_{i+1} + \ldots + p_{j-1})^2$.

The major upshot to scalar-scaffolded gluons is that, in these variables, the YM amplitude takes on a locked form, free from all redundancies. This is because scalar variables bake in momentum conservation and the on-shell conditions $\epsilon \cdot q = q^2 = 0$ from the start. We can compare this with the ambiguities present when the YM amplitude is written in the standard language, with polarization vectors and gluon momenta. There, for example, we can always use momentum conservation to solve for one momentum in terms of all the others $q_k^\mu = -\sum_{i \neq k} q_i^\mu$, and use this relation to get rid of all instances of $q_k^\mu$ in the amplitude. While we could simply make it a rule to always solve for one (arbitrary) momentum in terms of the others, this breaks manifest cyclic invariance of the object. By instead scalar-scaffolding the amplitude, we arrive at a canonical representation that keeps all symmetries intact. As an immediate corollary to these observations, in $X$ variables, acting with a gauge transformation $\epsilon_i^\mu \to \epsilon_i^\mu + \alpha_i q_i^\mu$ does absolutely \textit{nothing} to the form of the amplitude--- since there is only one way to write it. Instead, as we will review in the next section, gauge invariance (plus multi-linearity in polarization vectors) combine to produce a powerful constraint on the form of the amplitude.

As a result, when cast in scalar variables each individual term in the YM amplitude is well-defined. In other words, there is now meaning to the statement ``the numerator of the term with pole $X_{1,5}$...,'' whereas before this statement was plagued with ambiguity. Thus, the key insight here is to think of the YM amplitude as a Laurent series expansion in $X$ variables, giving us free rein to focus on understanding each term (or groups of terms) in the expansion individually, rather than needing to understand the whole object as one.  

In this letter, we leverage this critical advantage of the scalar variables to introduce and explore novel methods of computing YM tree-level amplitudes. With our new perspective, it is immediately clear one can use consistent gluon factorization to explicitly determine any term with poles in the $n$-point amplitude as a function of terms in lower-point amplitudes. Then, it turns out that \textit{gauge invariance} uniquely determines the term with no pole (the ``contact'' term): the $n$-point contact term can be computed uniquely from numerators of single $X$ variables in the $n$-point amplitude. While this indeed defines a recursion for the full amplitude, we can also use it to recurse only on terms in the amplitude with up to $k$ $X$'s in their denominators. Along the way, our construction explicitly proves that unitarity, locality, and gauge invariance uniquely determine YM tree amplitudes.

In the second part of this letter, we take inspiration from the computation of the contact term and construct a ``Laurent series recursion''--- where numerators in the $n$-point amplitude are determined solely by numerators in the same amplitude with one more pole. Taking as input \textit{only} the $(n-1)$-point amplitude (and no other lower-point amplitudes), we first use factorization to determine all cuts of the $n$-point amplitude for which $\mathcal{A}_n \to \mathcal{A}_3 \times \mathcal{A}_{n-1}$. From this initial set, we obtain all other terms recursively in the Laurent series expansion of the amplitude. And finally, based on the results of this Laurent series recursion, we derive a recursion just for the amplitude's contact term, fed only by the contact terms of lower-point amplitudes. 

Many of the findings presented in this letter complement previous uniqueness results for YM amplitudes \cite{Arkani-Hamed:2016rak,Rodina:2016mbk,Rodina:2016jyz,Rodina:2018pcb}. These recursion relations are also conceptually orthogonal to the ``standard'' on-shell recursion relations, which depend upon control on the amplitude as it approaches infinity in prescribed directions \cite{Arkani-Hamed:2008bsc,Britto:2004ap,Britto:2005fq,Arkani-Hamed:2024tzl}. As a happy consequence of this novelty, since here we are fundamentally computing amplitudes term-by-term (in arbitrary spacetime dimension and helicity configuration), our recursions return the amplitude in a manifestly local way, notably devoid of any spurious poles.

We include implementations of all three of these recursions in ancillary Mathematica files, which one can use to efficiently compute various terms in YM amplitudes. In particular, with \texttt{ContactTermRecursion.nb}, we derived the contact term of the $30$-point amplitude in just under $15$ minutes!

\vspace{\baselineskip}

\sect{Scalar-Scaffolding Review} Before proceeding to the results of this work, let us briefly review the scalar-scaffold formalism for gluon amplitudes \cite{Arkani-Hamed:2023jry}. To start, consider the tree-level scattering amplitude for $n$ gluons, which is a function of dot products of $n$ polarization vectors $\epsilon_i^\mu$ and gluon momenta $q_i^\mu$. We take this object and ``scaffold'' each gluon with two colored scalars via minimal coupling, so that the $n$-point gluon process is embedded \textit{inside} a $2n$-point scalar process. The gluon data is then mapped to the scalar data by $\epsilon_i^\mu = p_{2i}^\mu + \alpha_i (p_{2i-1}^\mu + p_{2i}^\mu)$ (for $\alpha_i$ an arbitrary gauge-dependent parameter) and $q_i^\mu = (p_{2i-1}^\mu + p_{2i}^\mu)$, and the amplitude becomes a function solely of scalar invariants $X_{i,j} = (p_i + p_{i+1} + \cdots + p_{j-1})^2$. By then localizing on the ``scaffolding residue'' (where each gluon is put on-shell), we land on the scalar-scaffolded $n$-point gluon amplitude $\mathcal{A}_n[X_{i,j}]$.

\begin{figure}[t]
    \centering
    \includegraphics[width=1.0\linewidth]{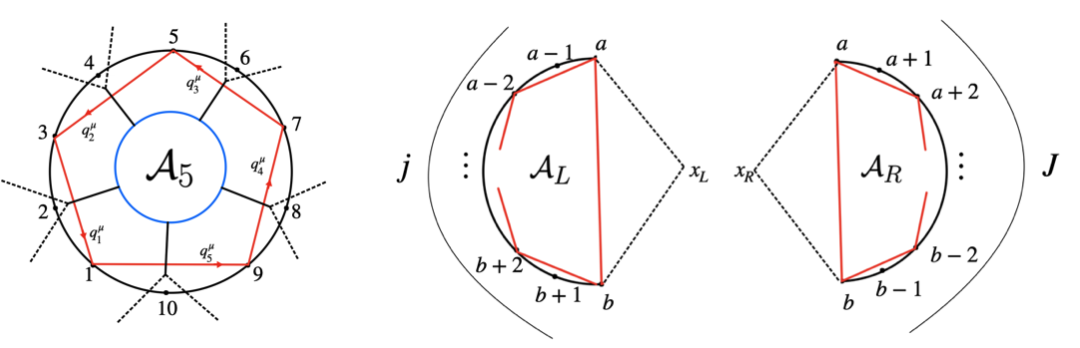}
    \caption{(Left) The five-point momentum polygon. Overlaid is the dual Feynman graph, where dotted lines represent scalars and solid black lines gluons. (Right) Gluon factorization on pole $X_{a,b} = 0$.}
    \label{fig:mom-poly-fact}
\end{figure}

Just as with amplitudes in Tr$(\phi^3)$ theory, the planar data  $X_{i,j}$ for the YM amplitude may be visualized as chords on the ``momentum polygon,'' shown (as a disk) at five-points in \cref{fig:mom-poly-fact} \footnote{In all figures, we draw momentum polygons as disks for readability and convenience.}. The momentum polygon is formed from $2n$ scalar dual coordinates $x_i^\mu$: gluons are then taken on-shell by enforcing $X_{2j - 1,2j + 1} = 0$ for all $j$, and the scalar on-shell conditions $X_{i,i+1} = 0$ imply transverse polarizations.

As mentioned in the Introduction, the YM amplitude has a completely unique form as a Laurent series in these $X$ variables: the scalar-scaffold formalism trivializes momentum conservation and the on-shell conditions, while maintaining manifest cyclic invariance. As such, in scalar variables, a gauge transformation $\epsilon_i^\mu \to \epsilon_i^\mu + \alpha_i q_i^\mu$ \textit{does} \textit{absolutely} \textit{nothing} to the amplitude. Instead, the requirement of gauge invariance in, say, gluon $k$ (as well as linearity in $\epsilon_k^\mu$) restricts the amplitude to be linear in each variable $X_{j,2k}$, and to satisfy the two constraints
\begin{equation}
\label{eq:scaf-gaug-inv}
    \mathcal{A}_n = \sum_{j \neq \{ 2k, 2k\pm1 \}} (X_{j,2k} - X_{j,2k \pm 1} ) \frac{\partial \mathcal{A}_n}{\partial X_{j,2k}},
\end{equation}
for all $k$. Additionally, we can translate into $X$'s the sum-over-polarizations that appears in gluon factorization channels: the residue of $\mathcal{A}_n$ on a gluon propagator $X_{a,b} = 0$ (with $a,b$ odd) $R_{a,b}$ may be written in scalar variables as
\begin{equation}
\label{eq:fact-form}
     R_{a,b} = \sum_{j \in L, J \in R} (X_{j,J} - X_{j,b} - X_{J,a}) \frac{\partial \mathcal{A}_L}{\partial X_{j,x_L}} \times \frac{\partial \mathcal{A}_R}{\partial X_{J,x_R}},
\end{equation}
where $L = \{b+1, b+2,\ldots, a-1 \}$, $R = \{a + 1, a + 2, \ldots, b-1\}$, and $\mathcal{A}_L$ and $\mathcal{A}_R$ are defined on the kinematics to the left and right, respectively, of the chord $X_{a,b}$; see \cref{fig:mom-poly-fact}.

\vspace{\baselineskip}

\sect{Factorization and Gauge Invariance} The scalar-scaffolded YM tree amplitude is a rational function of $X$ variables that may be canonically decomposed into a sum of terms, each characterized by its pole structure. Let us denote by $\mathcal{I} = \{ (i,j), (k,m), \ldots \}$ a set of non-intersecting chords on the momentum polygon, and by $N_\mathcal{I}$ the numerator of the term in $\mathcal{A}_n$ with denominator $X_\mathcal{I} \equiv X_{i,j} X_{k,m} \cdots$. Then, locality restricts the form of the amplitude to
\begin{equation}
\label{eq:A-gen-form}
    \mathcal{A}_n = N_0 + \sum_{\mathcal{I}} \frac{N_\mathcal{I}}{X_{\mathcal{I}}},
\end{equation}
with $N_0$ the ``contact term,'' and where the sum ranges over all possible sets $\mathcal{I}$ on the $2n$-point scaffolded momentum polygon.

\begin{figure}[t]
    \centering
    \includegraphics[width=0.7\linewidth]{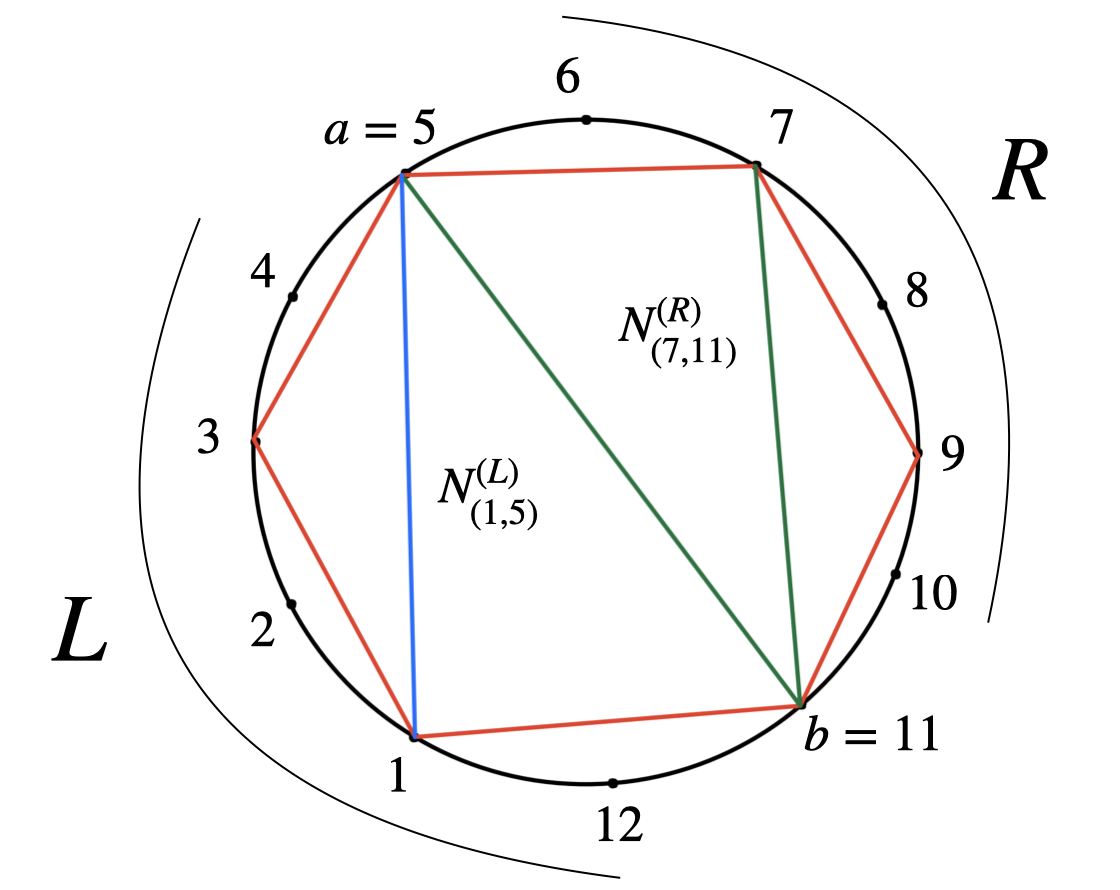}
    \caption{Consider computing $N_{\{ (5,11), (7,11) \} }$ (in green) by cutting on $X_{5,11}$ for the six-point amplitude using \cref{eq:N-fact}. Then, for the left surface, we have $\mathcal{I}_L = 0$ and $L_0 = \{ 1 \}$, while, for the right surface, $\mathcal{I}_R = \{ (7,11) \}$ and thus $R_0 = 0$. Therefore, we need to know the four-point numerators $N^{(L)}_0$, $N^{(L)}_{(1,5)}$ (in blue) and $N^{(R)}_{(7,11)}$.}
    \label{fig:fact-formula}
\end{figure}

For $\mathcal{I} \neq 0$, we can cut on any $(a,b) \in \mathcal{I}$, and the factorization formula~\eqref{eq:fact-form} tells us that
\begin{equation}
\begin{aligned}
\label{eq:N-fact}
    N_\mathcal{I} =& \sum_{j\in L,J\in R} ( X_{j,J} - X_{a,j} - X_{b,J} ) \frac{\partial N^{(L)}_{\mathcal{I}_L}}{\partial X_{x_L,j}} \frac{\partial N^{(R)}_{\mathcal{I}_R}}{\partial X_{x_R,J}} \\
    &- \left( \sum_{J \in R} \frac{\partial N^{(R)}_{\mathcal{I}_R}}{\partial X_{x_R,J}} \right) \left( \sum_{j \in L_o} \frac{\partial N^{(L)}_{\mathcal{I}_L\cup(a,j)}}{\partial X_{x_L,j}} \right) \\
    &- \left( \sum_{j \in L} \frac{\partial N^{(L)}_{\mathcal{I}_L}}{\partial X_{x_L,j}} \right) \left( \sum_{J \in R_o} \frac{\partial N^{(R)}_{\mathcal{I}_R\cup(b,J)}}{\partial X_{x_R,J}} \right).
\end{aligned}
\end{equation}
Here, $\mathcal{I}_L$ ($\mathcal{I}_R$) is the subset of $\mathcal{I}$ contained in the subsurface $L$ ($R$), with $\mathcal{I} = \mathcal{I}_L\cup\mathcal{I}_R\cup(a,b)$. Additionally, $L_o$ ($R_o$) is the set of odd indices in $L\cup\{ a,b \}$ ($R\cup\{ a,b \}$) for which $\mathcal{I}_{L}\cup(a,j)$ for $j \in L_o$ ($\mathcal{I}_{R}\cup(b,J)$ for $J \in R_o$) describes a local pole. In \cref{fig:fact-formula}, we give an visual example of how to use this formula to compute a numerator in the six-point amplitude.

This determines all terms with poles, but what about when $\mathcal{I} = 0$? The scaffolding statement of gauge invariance~\eqref{eq:scaf-gaug-inv} in the $n^{\rm{th}}$ gluon dictates that $N_0$ must satisfy the following identity:
\begin{equation}
\label{eq:contact-gaug-form}
    N_0 = \sum_{j = 2}^{2n-2} (X_{j,2n} - X_{j,2n-1} ) \frac{\partial N_0}{\partial X_{j,2n}} - \sum_{j \in R} \frac{\partial N_{(j,2n-1)}}{\partial X_{j,2n}},
\end{equation}
where $R = \{ 3, 5, \ldots, 2n - 5\}$. Since $N_0$ is $O(X^2)$, we can write it generally as
\begin{equation}
    N_0 = \sum_{a,b,c,d} c_{(a,b);(c,d)} X_{a,b} X_{c,d},
\end{equation}
where $a,b,c,d$ range over all points on the momentum polygon. Thus, in order to determine $N_0$, we are tasked with fixing all $c$'s in the above expansion.

\begin{figure}[t]
    \centering
    \includegraphics[width=0.9\linewidth]{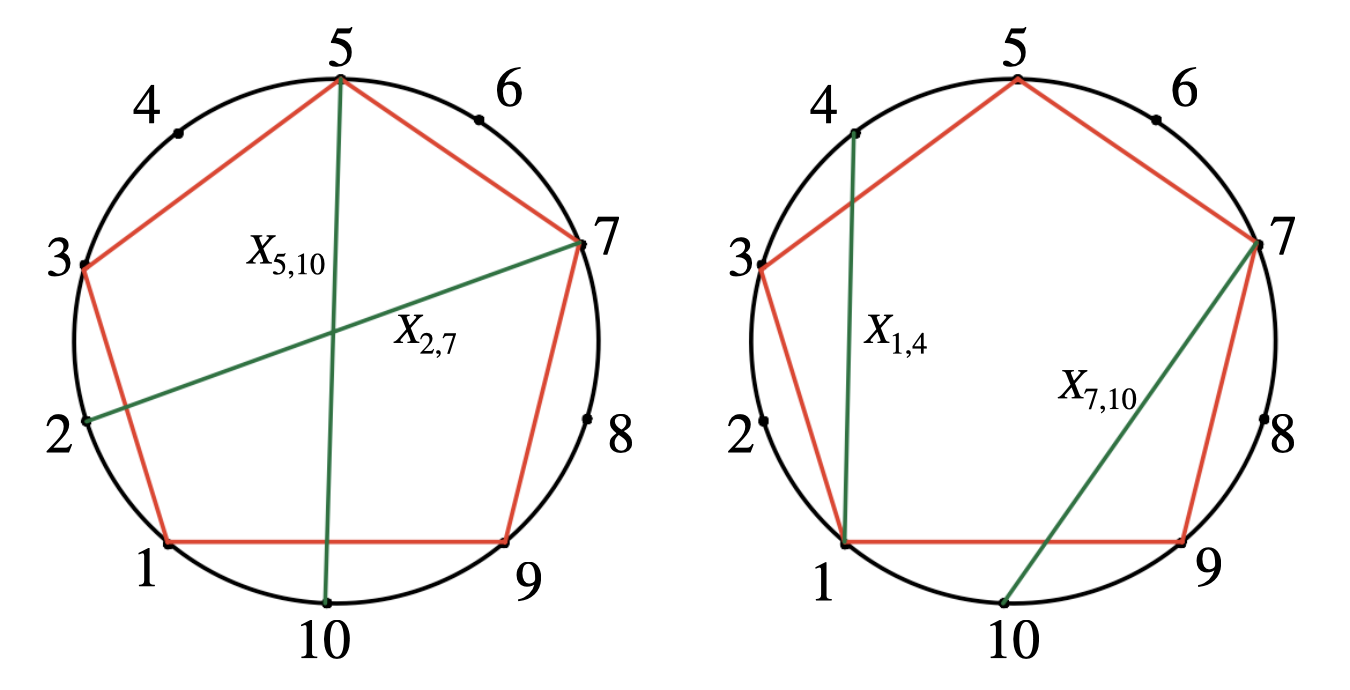}
    \caption{At five-points, any two chords drawn on the momentum polygon always leave at least two neighboring points open: $e.g.$, points $3,4$ on the left, and points $5,6$ on the right.}
    \label{fig:open-pts}
\end{figure}

Let us start by asking for all $c_{(a,b);(c,d)}$ where $a,b,c,d \neq 2n-1,2n$. In this case, we can take two derivatives of \cref{eq:contact-gaug-form} with respect to $X_{a,b}$ and $X_{c,d}$. These derivatives pass right through the $(X_{j,2n} - X_{j,2n-1})$ prefactor and kill $\partial_{X_{j,2n}} N_0$ purely by units. We are therefore left with
\begin{equation}
\label{eq:c-det-2n}
    c_{(a,b);(c,d)} = - \sum_{j \in R} \frac{\partial^3 N_{(j,2n-1)}}{\partial X_{a,b} \partial X_{c,d} \partial X_{j,2n}}.
\end{equation}
That is, gauge invariance requires that these $c$'s are uniquely fixed by terms in $\mathcal{A}_n$ with exactly one pole! To determine the other $c$'s, we must use gauge invariance in the other gluons. From the argument given above, it is clear that we can perform the same trick for any $c_{(a,b);(c,d)}$ as long as the chords $X_{a,b}$ and $X_{c,d}$ leave a space of at least two neighboring points open as drawn on the momentum polygon; we give examples of this at five-points in \cref{fig:open-pts}. One can quickly verify that this is always true for amplitudes with multiplicity $n \geq 5$, so the contact piece for these amplitudes is readily fixed by formulae like \cref{eq:c-det-2n} \footnote{For $n = 4$, only $c_{(1,5);(3,7)}$, $c_{(2,6);(4,8)}$, $c_{(2,8);(4,6)}$, and $c_{(2,4);(6,8)}$ do not satisfy the aforementioned criterion, and thus may not be determined by this simple trick. But, in the case at four-points, it is well-appreciated that gauge invariance is responsible for the contact term.}.

So, in scaffolding language, gauge invariance delivers for us an explicit and unique determination of the contact piece $N_0$ of $\mathcal{A}_n$ as a function solely of numerators of terms in $\mathcal{A}_n$ with one pole \footnote{One may hope that this procedure can determine \textit{any} (non-leading-singularity) numerator as a function of numerators with one more pole. However, for this simple trick to work fully for the numerator of a term with $p$ poles, one requires a $(2p + 5)$-point amplitude. The same counting holds when we add $F^3$ and higher irrelevant terms to YM: in this case, equations analogous to \cref{eq:c-det-2n} fully determine the contact term at $O(X^m)$ for amplitudes at $2m+1$ points and higher.}. Therefore, we have demonstrated --- by way of an algorithm for doing so --- that locality, unitarity, and gauge invariance uniquely fix $\mathcal{A}_n$.

Of course, this construction is also a valid recursion for the amplitude. In fact, looking closer at \cref{eq:N-fact}, we see that $N_\mathcal{I}$ depends only on lower-point numerators with at most as many poles as $\mathcal{I}$. Since the contact term only requires knowledge of single-$X$ numerators, this means that we can recurse not only on the full amplitude but on all terms up to \textit{any} order in the Laurent series expansion! In one attached Mathematica file \texttt{FirstTwoRecursion.nb}, we implement this recursive procedure for the first two orders, the contact term and single-$X$ terms.

In the next section of this letter, we will show --- using a bit more machinery --- that it is indeed possible to define a ``Laurent series recursion'' for the amplitude, where numerators are written as functions exclusively of numerators with one more pole.

\vspace{\baselineskip}

\begin{figure}[t]
    \centering
    \includegraphics[width=1.0\linewidth]{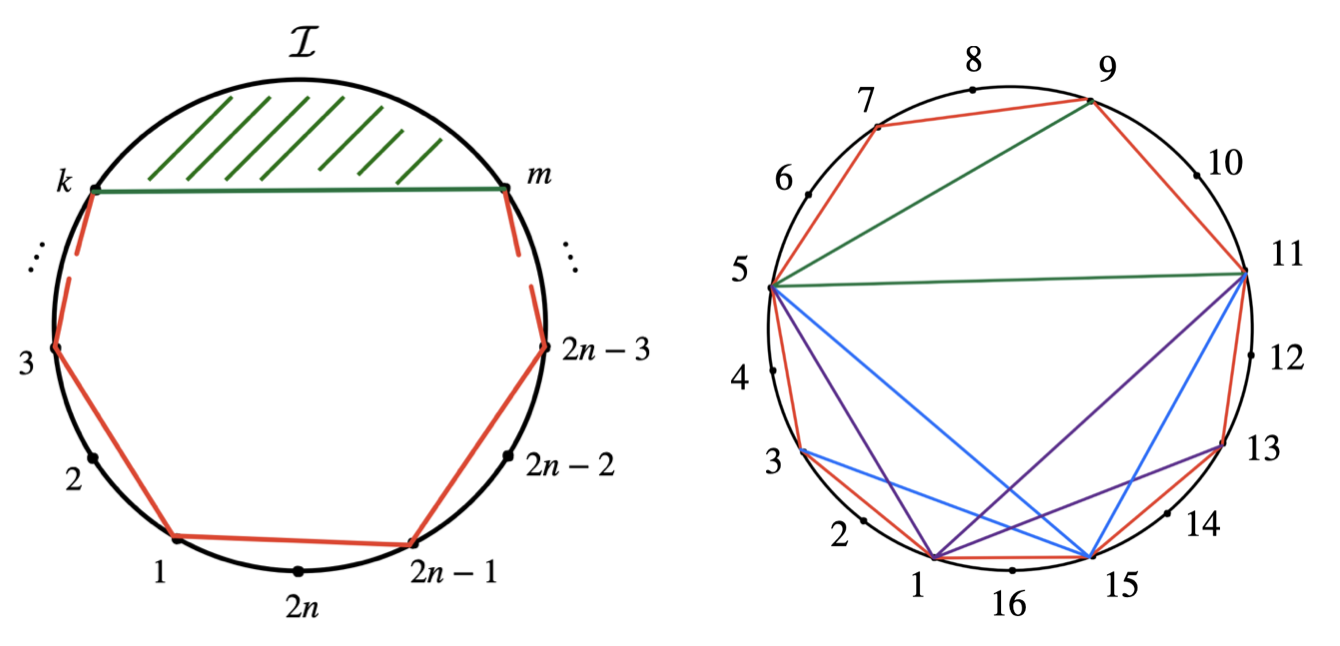}
    \caption{(Left) The set $\mathcal{I}$ contains the chord $X_{k,m}$ for $k,m \neq 1, 2n-1$ as well as any collection of non-intersecting chords in the shaded green region ``behind'' it. (Right) To determine $N_{(5,11),(5,9)}$ of the eight-point amplitude, we need to know $T_{1}^{(16)}$ (purple and green), $T_{15}^{(16)}$ (blue and green), as well as $a_{i,j}$ (purple chord $X_{1,5}$ and green).}
    \label{fig:set-I-Laur}
\end{figure}

\sect{Laurent Series Recursion} Let us now use the requirement of gauge invariance to constrain terms with poles in $\mathcal{A}_n$. Without loss of generality, we will focus on the constraint coming from gauge invariance in the $n^{\rm{th}}$ gluon. We are then interested in numerators $N_\mathcal{I}$ for sets $\mathcal{I}$ for which all chords lie ``behind'' some chord $X_{k,m}$ for $k,m \neq 1, 2n-1$, as shown on the $l.h.s.$ of \cref{fig:set-I-Laur}. For numerators satisfying these properties (including $N_0$), the scaffolding statements of gauge invariance read
\begin{equation}
\begin{gathered}
\label{eq:N-gauge-inv}
    N_\mathcal{I} = \sum_{j = 2}^{2n-2} ( X_{j,2n} - X_{j,1} ) \frac{\partial N_\mathcal{I}}{\partial X_{j,2n}} - T^{(2n)}_1, \\
    N_\mathcal{I} = \sum_{j = 2}^{2n-2} ( X_{j,2n} - X_{j,2n-1} ) \frac{\partial N_\mathcal{I}}{\partial X_{j,2n}} - T^{(2n)}_{2n-1},
\end{gathered}
\end{equation}
where, like for the contact term, the $T$'s are composed of numerators in $\mathcal{A}_n$ with just one more $X$ in their denominators:
\begin{equation}
\begin{gathered}
\label{eq:T-defs}
    T^{(2n)}_1 \equiv \sum_{j \in R_1} \frac{ \partial N_{\mathcal{I}\cup(j,1)}}{\partial X_{j,2n}}, \\
    T^{(2n)}_{2n-1} \equiv \sum_{j \in R_2} \frac{\partial N_{\mathcal{I}\cup(j,2n-1)}}{\partial X_{j,2n}},
\end{gathered}
\end{equation}
with sets $R_1 = \{ 5, 7, \ldots, k, m, \ldots, 2n-3\}$ and $R_2$ = $\{ 3, 5, \ldots, k, m, \ldots, 2n-5\}$. Subtracting the two representations in \cref{eq:N-gauge-inv} then yields the following constraint:
\begin{equation}
\label{eq:gaug-inv-const}
    \Delta T_{2n} = \sum_{j = 2}^{2n-2} \delta_j \left( \frac{\partial N_\mathcal{I}}{\partial X_{j,2n}} \right),
\end{equation}
where $\Delta T_{2n} \equiv T_1^{(2n)} - T_{2n-1}^{(2n)}$ and $\delta_j \equiv X_{j,2n-1} - X_{1,j}$ are the ``soft factors,'' or the quantities that vanish as we take the $n^{\rm{th}}$ gluon soft. Our object in this section is to solve \cref{eq:gaug-inv-const} (and other gauge-invariance constraints) and derive expressions for the $\partial_{X_{j,2n}} N_\mathcal{I}$ depending only on terms with one more pole in their denominators.

Now, taking a closer look at \cref{eq:gaug-inv-const}, it must be that we can decompose $\Delta T_{2n}$ as
\begin{equation}
\label{eq:Delta-T-decomp}
    \Delta T_{2n} = \sum_{j = 2}^{2n-2} \delta_j F_j,
\end{equation}
where the $F_j$ are polynomials which can, generically, depend on any $\delta_i$, and therefore are non-unique. This is the only possibility because, by construction, neither side of \cref{eq:gaug-inv-const} has poles.

As our first step towards deriving the $\partial_{X_{j,2n}} N_\mathcal{I}$, let us find a set of $F_j$ that solves \cref{eq:Delta-T-decomp}. This turns out to have a straightforward solution: in App. A, we show that $\Delta T_{2n}$ is actually a linear form in the $\delta_j$, and thus there exist simple expressions for the $F_j$. In particular, we find
\begin{equation}
    F_i = \frac{\partial \Delta T_{2n}}{\partial \delta_i},
\end{equation}
for $i = 2, 3, 2n-3, 2n-2$, and
\begin{equation}
    F_i = \frac{\partial \Delta T_{2n}}{\partial X_{i,2n-1}}\bigg \vert_{\mathcal{S}_1} = -\frac{\partial \Delta T_{2n}}{\partial X_{1,i}}\bigg \vert_{\mathcal{S}_2},
\end{equation}
for $i = 4, 5, \ldots, 2n-4$, where $\mathcal{S}_1$ instructs us to take $\delta_{2n-3}, \delta_{2n-2} \to 0$, and $\mathcal{S}_2$ the same for $\delta_2, \delta_3$. 

However, these $F_i$ do not provide unique solutions to \cref{eq:gaug-inv-const}. In fact, any expression of the form
\begin{equation}
\label{eq:NI-form}
    \frac{\partial N_\mathcal{I}}{\partial X_{i,2n}} = F_i + a_{i,j} \delta_j + a_{i,j,k} \delta_j \delta_k + \cdots
\end{equation}
will solve \cref{eq:gaug-inv-const}, so long as the tensors $a$ satisfy certain conditions; for example, $a_{i,j}$ is required to be a skew-symmetric matrix. Therefore, if we want to derive a form for the $\partial_{X_{i,2n}} N_\mathcal{I}$, we will need to use other constraints in order to shed light on the $a$ tensors. Again, gauge invariance provides the solution, just now in the neighboring gluons.

As we show in App. B, all $a$ tensors with three or greater indices vanish, and $a_{i,j}$ only has eight non-zero entries. Putting everything together, with the form of $N_\mathcal{I}$ shown in \cref{eq:N-gauge-inv}, we find
\begin{equation}
\label{eq:final-recur-1}
    \frac{\partial N_\mathcal{I}}{\partial X_{j,2n}} = \frac{\partial \Delta T_{2n}}{\partial \delta_j} \bigg \vert_{\mathcal{S}} + a_{i,j} \delta_j,
\end{equation}
where $\mathcal{S}$ does nothing for $j = 2, 3, 2n-3, 2n-2$, while, for $j = 4, 5, \ldots, 2n-4$, it instructs us to rewrite $X_{i,2n-1} = X_{i,1} + \delta_i$ and take $\delta_{2n-3},\delta_{2n-2} \to 0$. The matrix $a_{i,j}$ is skew-symmetric, and all of its elements except
\begin{equation}
\label{eq:final-recur-2}
    a_{i,2} = \frac{\partial^3 N_{\mathcal{I}\cup(1,l)}}{\partial X_{2,l} \partial X_{i,2n} \partial X_{3,2n-1}}, \quad a_{i,3} = -a_{i,2},
\end{equation}
for $i = 2n-3, 2n-2$ (and those related by skew-symmetry) vanish. In the above, $l = m$ when $k = 3$ and $l = 5$ otherwise. Of course, we can permute these equations, so there is one for each external gluon in the amplitude. Both \cref{eq:final-recur-1,eq:final-recur-2} compute $N_\mathcal{I}$ as a function exclusively of terms in $\mathcal{A}_n$ with one more $X$ in their denominator, thus achieving a recursion in the Laurent series expansion of the amplitude. An example of this construction is given at eight-points on the $r.h.s.$ of \cref{fig:set-I-Laur}.

In \texttt{LaurentRecursion.nb}, we give a particular realization of this Laurent series recursion for the full amplitude. We take as input all cuts of $\mathcal{A}_5$ where $\mathcal{A}_5 \to \mathcal{A}_3 \times \mathcal{A}_4$. We then use \cref{eq:final-recur-1,eq:final-recur-2} to recursively compute all other terms in $\mathcal{A}_5$ down to the contact term. From this, we determine all cuts of $\mathcal{A}_6$ where $\mathcal{A}_6 \to \mathcal{A}_3 \times \mathcal{A}_5$, and thus the cycle continues up to the desired multiplicity. In addition to returning the full amplitude, the implementation can also return a list that contains each of its individual terms. 

Note that, although we input full knowledge of the lower-point amplitude, the recursion relations do not require all its terms. For example, $N_{(1,5)}$ for the five-point amplitude may be given initially by factorization into $\mathcal{A}_4 \times \mathcal{A}_3$, or it can be determined using \cref{eq:final-recur-1,eq:final-recur-2} with knowledge of $N_{\{(1,5),(5,9)\}}$ and $N_{\{(1,5),(1,7)\}}$.

\vspace{\baselineskip}

\sect{Contact Term Recursion} For both of the recursions we have explored in this letter, an amplitude's contact term is determined (via gauge invariance) from its single-$X$ cuts. If we wanted to compute these cuts, we could use the factorization formula~\eqref{eq:N-fact}, where each single-$X$ cut is given as a function of lower-point contact terms and single-$X$ cuts. Ultimately, therefore, single-$X$ cuts are determinable from lower-point contact terms. It is then natural to ask if there exists a recursion for the contact term which relies on the contact terms of lower-point amplitudes and nothing more.

In App. F, we write down a set of recursive equations that do precisely this. They follow directly from the factorization formula for single-$X$ cuts and \cref{eq:final-recur-1,eq:final-recur-2} for $\mathcal{I} = 0$. We give an implementation of this recursion in \texttt{ContactRecursion.nb}, where, $e.g.$, we were able to compute the contact term of the 30-point amplitude in under $15$ minutes!

\vspace{\baselineskip}

\sect{Outlook} The scalar-scaffold formalism recasts the YM amplitude as a well-defined Laurent series expansion in $X$ variables. In this letter, taking up this perspective, we introduced several novel methods of computing (parts of) YM tree-level amplitudes. 

We began by writing down the most obvious term-by-term recursion, by using locality and consistent factorization on poles ($i.e.$, unitarity), which holds for any term with poles. For the remaining contact term, the scalar-scaffolded statement of gauge invariance provided us with a straightforward means of computing it, from terms with just one $X$ in their denominators. Surprisingly, this allowed us not only to define a closed recursion for the full amplitude but also for all terms up to any order in the Laurent series. We then demonstrated that, starting from an initial set of high-order terms in the Laurent series, we could recursively determine all other terms in the amplitude down to the contact term. Since this initial set depended only on the $(n - 1)$-point amplitude, this procedure also defines a recursion across multiplicity, where, interestingly, the $n$-point amplitude relies \textit{only} on the $(n-1)$-point amplitude. Lastly, we wrote down a recursion directly for the contact term of the amplitude; a simple implementation of this algorithm can generate high-point contact terms quickly.

Let us now discuss some of the many possible directions for future work.

Our Laurent series recursion for the amplitude begins with knowledge of (a subset of) cuts on which $\mathcal{A}_n \to \mathcal{A}_3 \times \mathcal{A}_{n-1}$. Generally, this subset contains more than just the cuts with the highest codimension ($i.e.$, leading singularities). At eight-points, for example, the numerator with pole $X_{1,5}X_{5,9}X_{9,13}X_{1,13}$ cannot be derived from our methods and must be part of the input--- even though it does not correspond to a leading singularity. It is then natural to ask if there exist additional recursive relations that allow us to construct the entire amplitude just from knowledge of its leading singularities. 

It seems likely that the contact term recursion given in App. F can be directly \textit{solved}, which would land us on a formula for any contact term as a function of three-particle amplitudes and soft factors. Just as this contact term recursion was derivable from our Laurent series recursion, so too do we expect analogous formulae for \textit{any} term in the amplitude expansion to follow. Might it then be possible to solve all recursions and write down a term-by-term formula for the full scalar-scaffolded YM amplitude, valid in all dimensions and for all helicity configurations? Could this be pathway to understanding the YM amplitude as one takes the multiplicity $n$ to be large? 

The ``nice'' properties of the scalar-scaffolded YM tree amplitude --- the scaffolded statements of gauge invariance and factorization  --- that led to the results in this letter all have natural generalizations to the one-loop ``surface integrand,'' defined on an extended basis of kinematics known as ``surface kinematics'' \cite{Arkani-Hamed:2024tzl}. Of course, it would be fasciniating to extend these ideas and recursions to the surface integrand.

\vspace{\baselineskip}

\sect{Acknowledgments} I thank Nima Arkani-Hamed, Carolina Figueiredo, and Jared Goldberg for inspiring discussions. This work was supported by the NSF Graduate Research Fellowship under Grant No. KB0013612. 

\appendix
\onecolumngrid

\section{Appendix A: $\Delta T_{2n}$ is linear in $\delta_j$}

In this Appendix, we show that the quantity $\Delta T_{2n}$ is a linear form in the soft factors $\delta_j$, and we give a solution for the $F_j$ in \cref{eq:Delta-T-decomp}. 

To start, we must demonstrate that (1) $T_1^{(2n)}$ depends at most linearly on the indices $1$ and $2n-1$, and (2) the only time $1$ and $2n-1$ can appear together is when either $X_{2,2n-1}$ or $X_{3,2n-1}$ multiplies $X_{1,p}$ for $p = 4, 5, \ldots, 2n-4$. We proceed via induction. We can verify this explicitly for the four-point amplitude, where $T_1^{(8)} = \partial_{X_{5,8}} N_{(1,5)}$ satisfies these properties.

Let us consider a set of chords $\mathcal{I}$ satisfying the properties outlined in the main text. For any odd $i$ (not between $k$ and $m$), we can cut $\mathcal{A}_n$ on the pole $X_{i,1} = 0$, and factorization~\eqref{eq:N-fact} yields the following formula:
\begin{equation}
\begin{aligned}
\label{eq:fact-N-1}
    \frac{\partial N_{\mathcal{I}\cup(i,1)}}{\partial X_{i,2n}} = \sum_{j\in R, J\in L} (X_{j,J} - X_{j,1} - X_{J,i}) &\frac{\partial^2 N_\mathcal{J}^{(R)}}{\partial X_{i,2n} \partial X_{x_R, j}} \frac{\partial N_\mathcal{K}^{(L)}}{\partial X_{x_L,J}} - \left( \sum_{j} \frac{\partial^2 N_\mathcal{J}^{(R)}}{\partial X_{i,2n} \partial X_{x_R, j}} \right) \left( \sum_J \frac{\partial N_{\mathcal{K}\cup (J,i)}^{(L)}}{\partial X_{x_L,J}} \right) \\
    &- \left( \sum_{j} \frac{\partial^2 N_{\mathcal{J}\cup(j,1)}^{(R)}}{\partial X_{i,2n} \partial X_{x_R, j}} \right) \left( \sum_J \frac{\partial N_{\mathcal{K}}^{(L)}}{\partial X_{x_L,J}} \right),
\end{aligned}
\end{equation}
where $\mathcal{J} = 0, \mathcal{K} = \mathcal{I}$ if $i \geq m$, and $\mathcal{J} = \mathcal{I}, \mathcal{K} = 0$ if $i \leq k$. Further, we have $L = \{ 2, 3, \ldots, i - 1\}$ and $R = \{ i + 1, i + 2, \ldots, 2n \}$.

Throughout this derivation, in addition to the inductive hypothesis, we will require three other facts: (1) the double derivative of $N_\mathcal{J}^{(R)}$ in the first line is independent of both $1$ and $2n-1$, (2) any numerator $N_\mathcal{I}$ where $\mathcal{I}$ does not touch the index $i$ is at most linear in $i$, and (3) $\partial_{X_{x_L,J}}N_\mathcal{K}^{(L)}$ is independent of $1$ unless $J = 2, 3$. We prove these claims in App. C.

So, assuming these facts, let us work through \cref{eq:fact-N-1} line-by-line. Dependence on $2n-1$ can only enter the first line linearly through the prefactor of the first term, when $j = 2n-1$. The only place where the first line could be \textit{quadratic} in $1$ is with the combination $X_{j,1} \mathcal{W}_{x_L}[N_{\mathcal{K}}^{(L)}]$, where
\begin{equation}
    \mathcal{W}_{x_L}[N_{\mathcal{K}}^{(L)}] \equiv \sum_{J} \frac{\partial N_\mathcal{K}^{(L)}}{\partial X_{x_L,J}}.
\end{equation}
This operator was introduced in Ref.~\cite{Backus:2025njt}, where it was shown to obey a set of sum rules in powers of soft factors. Since we know that $N_\mathcal{K}^{(L)}$ is at most linear in $1$, these sum rules imply that $W_{x_L}[ N_\mathcal{K}^{(L)}]$ is a sum of derivatives with respect to $1$. Thus, the first line of \cref{eq:fact-N-1} is linear in $1$ and $2n-1$.

The only place where $2n-1$ and $1$ may appear together in the first line is when $j = 2n-1$, where we find
\begin{equation}
    X_{J,2n-1} \mathcal{W}_{x_R}\left[ \frac{\partial N_{\mathcal{J}}^{(R)}}{\partial X_{i,2n}}\right] \frac{\partial N_\mathcal{K}^{(L)}}{\partial X_{x_L,J}} 
\end{equation}
The middle term is independent of $1$, whereas the last term only depends (linearly) on $1$ when $J = 2,3$. Hence, we find that $X_{2,2n-1} X_{1,p}$ and $X_{3,2n-1} X_{1,p}$ for $p = 4, 5, \ldots, 2n-4$ can appear in \cref{eq:fact-N-1}.

\begin{figure}[t]
    \centering
    \includegraphics[width=0.6\linewidth]{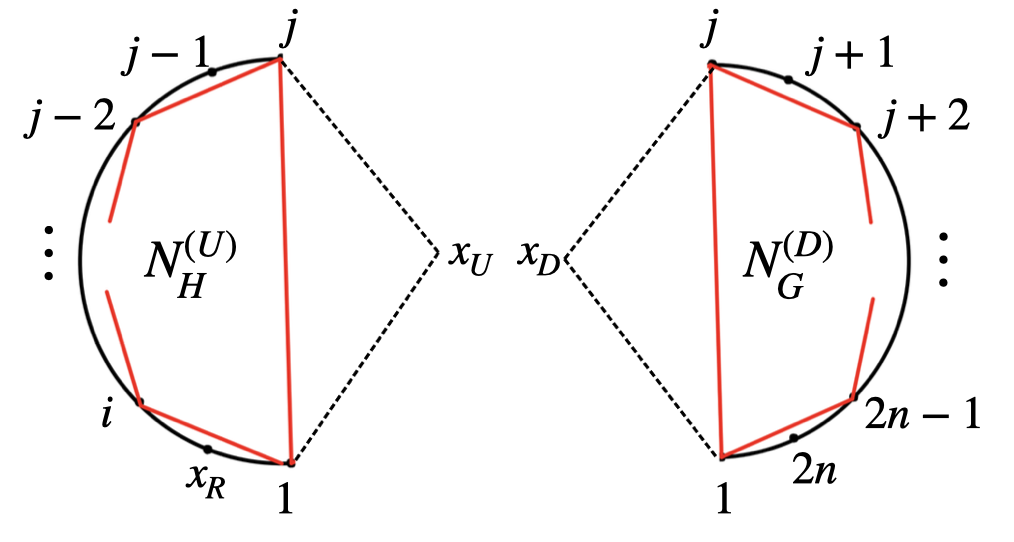}
    \caption{The factorization of $ N_{\mathcal{J}\cup(j,1)}^{(R)}$ in \cref{eq:for-fig-6} on cut $X_{j,1} = 0$.}
    \label{fig:fact-on-R}
\end{figure}

Finally, in the second line, the second term is clearly independent of both $1$ and $2n-1$. The first term is slightly trickier. When $\mathcal{J} = 0$, it is $O(X)$ by units, and thus it is at most linear in $1$ and $2n-1$. When $\mathcal{J} = \mathcal{I}$, for any $j$ we can cut on $X_{j,1} = 0$ and apply the derivatives. This yields
\begin{equation}
\label{eq:for-fig-6}
    \frac{\partial^2 N_{\mathcal{J}\cup(j,1)}^{(R)}}{\partial X_{i,2n} \partial X_{x_R, j}} = \left( \frac{\partial^2 N_\mathcal{H}^{(U)}}{\partial X_{x_R,j} \partial X_{i,x_U}} \right) \left( \frac{\partial N_{\mathcal{G}}^{(D)}}{\partial X_{x_D,2n}} \right),
\end{equation}
where $\mathcal{H} = 0, \mathcal{G} = \mathcal{I}$ when $j \leq k$ and $\mathcal{H} = \mathcal{I}, \mathcal{G} = 0$ when $j \geq m$. (In \cref{fig:fact-on-R}, we show this factorization of $ N_{\mathcal{J}\cup(j,1)}^{(R)}$ graphically.) In the first case, the first term is a pure number by units, and the second term is linear in $1$ and independent of $2n-1$. In the second case, the first term depends on neither $1$ nor $2n-1$ due to the proof in App. C, and the second term is $O(X)$. Thus, we have completed the proof.

Of course, an identical argument can show the analogous for $T_{2n-1}^{(2n)}$, where now it is $\delta_{2n-3}, \delta_{2n-2}$ that may appear with any $X_{p,2n-1}$ for $p = 4, 5, \ldots, 2n-4$. These results imply that $\Delta T_{2n}$ is a linear form in the $\delta_j$, as well as give a simple solution for the $F_j$ appearing in \cref{eq:Delta-T-decomp}: we find
\begin{equation}
    F_i = \frac{\partial \Delta T_{2n}}{\partial \delta_i}
\end{equation}
for $i = 2, 3, 2n-3, 2n-2$. For $i = 4, 5, \ldots, 2n-4$, we must be a bit more careful due to the small mixing between indices $1$ and $2n-1$. This can, however, be easily dealt with by writing
\begin{equation}
    F_i = \frac{\partial \Delta T_{2n}}{\partial X_{i,2n-1}}\bigg \vert_{\mathcal{S}_1} = -\frac{\partial \Delta T_{2n}}{\partial X_{1,i}}\bigg \vert_{\mathcal{S}_2},
\end{equation}
where $\mathcal{S}_1$ instructs us to take $\delta_{2n-3}, \delta_{2n-2} \to 0$, and $\mathcal{S}_2$ the same for $\delta_2, \delta_3$. Thus, all told, we conclude that we may express $\Delta T_{2n}$ as
\begin{equation}
\begin{aligned}
    \Delta T_{2n} = \sum_{j = 2,3, 2n-3,2n-2} &\delta_j \left( \frac{\partial \Delta T_{2n}}{\partial \delta_j} \right) \\
    &+ \sum_{j = 4}^{2n-4}\delta_j \left( \frac{\partial \Delta T_{2n}}{\partial X_{j,2n-1}} \right)_{\mathcal{S}_1},
\end{aligned}
\end{equation}
where the coefficient of $\delta_i$ is independent of $1$ for $i = 4, 5, \ldots, 2n-2$ and independent of $2n-1$ for $i = 2, 3, \ldots, 2n-4$.

\section{Appendix B: $a$ tensors from gauge invariance in adjacent gluons}

Here, we solve for the $a$ tensors appearing in \cref{eq:NI-form} by imposing gauge invariance in gluon $1$ and gluon $n-1$. Let us begin with gluon $1$, whose gauge invariance implies that
\begin{equation}
\begin{aligned}
\label{eq:gaug-inv-1}
    \frac{\partial N_\mathcal{I}}{\partial X_{i,2n}} &= \sum_{j = 4}^{2n} ( X_{j,2} - X_{j,3} ) \frac{\partial^2 N_\mathcal{I}}{\partial X_{i,2n} \partial X_{j,2}} - \frac{\partial T^{(2)}_3 }{\partial X_{i,2n}} \\
    &= F_i + a_{i,j} \delta_j + a_{i,j,k} \delta_j \delta_k + \cdots
\end{aligned}
\end{equation}
for $i = 4, 5, \ldots, 2n-2$, where of course
\begin{equation}
\label{eq:T-3-2-def}
    \frac{\partial T^{(2)}_3 }{\partial X_{i,2n}} = \sum_j \frac{\partial^2 N_{\mathcal{I}\cup(j,3)}}{\partial X_{i,2n} \partial X_{2,j}}.
\end{equation}
From the results of App. C, we conclude that \cref{eq:gaug-inv-1} is independent of the index $1$ for $i = 4, 5, \ldots, 2n-2$. Using gauge invariance in gluon $n-1$, the analogous argument tells us that $\partial_{X_{i,2n}} N_\mathcal{I}$ is independent of index $2n-1$ for $i = 2, 3, \ldots, 2n-4$. And, since we proved (in App. A) that these properties are also those of the $F_i$, we are forced to conclude that all $a_{i,j,k,\cdots} = 0$ whenever $i = 4, 5, \ldots, 2n-4$. That is, using gauge invariance in the adjacent two gluons, we have demonstrated that
\begin{equation}
     \frac{\partial N_\mathcal{I}}{\partial X_{i,2n}} = \frac{\partial \Delta T_{2n}}{\partial X_{i,2n-1}}\bigg \vert_{\mathcal{S}_1} = -\frac{\partial \Delta T_{2n}}{\partial X_{1,i}}\bigg \vert_{\mathcal{S}_2},
\end{equation}
for $i = 4, 5, \ldots, 2n-4$. It also tells us that, when $i = 2n-3,2n-2$, the only $\delta$'s that can appear in the expansion~\eqref{eq:NI-form} are $\delta_2, \delta_3$; and, for $i = 2,3$, only $\delta_{2n-3}, \delta_{2n-2}$.

Let us first examine the case when $i = 2n-3, 2n-2$ and look for instances of $\delta_2,\delta_3$ in \cref{eq:NI-form}. The derivative of $T_3^{(2)}$ can never depend on $\delta_2 = X_{2,2n-1}$ by linearity in gluon 1, and it also can never depend on $\delta_3 = X_{3,2n-1}$ since it must be independent of $3$ by the results of App. A. Additionally, the results of App. C tell us that $\partial_{X_{i,2n}} \partial_{X_{j,2}} N_\mathcal{I}$ does not depend on $2n-1$ for $i = 2n-3,2n-2$. This means that, for these values of $i$,
\begin{equation}
\label{eq:doub-der}
    a_{i,j} \delta_j + a_{i,j,k} \delta_j \delta_k + \cdots = (\delta_2 - \delta_3) \frac{\partial^2 N_\mathcal{I}}{\partial X_{i,2n} \partial X_{2,2n-1}}.
\end{equation}
So, we now only need to work out a formula for $\partial_{X_{i,2n}} \partial_{X_{2,2n-1}} N_\mathcal{I}$. To do this, we note that gauge invariance in gluon $1$ (plus all of the preceding arguments) tells us that
\begin{equation}
    \frac{\partial N_\mathcal{I}}{\partial X_{2,2n-1}} = \frac{\partial \Delta T_2}{\partial \overline{\delta}_{2n-1}} + (\overline{\delta}_4 - \overline{\delta}_5) \frac{\partial^2 N_\mathcal{I}}{\partial X_{2,2n-1} \partial X_{1,4}},
\end{equation}
for all $\mathcal{I}$ with $k \neq 3$, where $\overline{\delta}_i \equiv X_{1,i} - X_{3,i}$ are the soft factors for gluon $1$. From App. A, the derivative of $\Delta T_2$ cannot depend on either $\overline{\delta}_4,\overline{\delta}_5$. Since we know from App. C that $\partial_{X_{i,2n}} \partial_{X_{2,2n-1}} N_\mathcal{I}$ does not depend on $1$, we necessarily have
\begin{equation}
\label{eq:app-D-gen}
    \frac{\partial^2 N_\mathcal{I}}{\partial X_{i,2n} \partial X_{2,2n-1}} = \frac{\partial^2 \Delta T_2}{\partial X_{i,2n} \partial \overline{\delta}_{2n-1}}.
\end{equation}
Hence, we find that the only nonzero $a$ tensor elements are
\begin{equation}
\label{eq:lower-left-indices}
    a_{i,2} = \frac{\partial^2 \Delta T_2}{\partial X_{i,2n} \partial \overline{\delta}_{2n-1}}, \quad a_{i,3} = -\frac{\partial^2 \Delta T_2}{\partial X_{i,2n} \partial \overline{\delta}_{2n-1}}
\end{equation}
for $i = 2n-3,2n-2$. In App. E, we show that the above also holds for $k = 3$. 

Now, we can go through the analogous argument using gauge invariance in gluon $n - 1$, where we will predictably discover, for $i = 2,3$, that
\begin{equation}
\label{eq:upper-right-indices}
    a_{i,2n-2} = -\frac{\partial^2 \Delta T_{2n-2}}{\partial X_{i,2n} \partial \tilde{\delta}_1}, \quad a_{i,2n-3} = \frac{\partial^2 \Delta T_{2n-2}}{\partial X_{i,2n} \partial \tilde{\delta}_1}
\end{equation}
where $\tilde{\delta}_j \equiv X_{j,2n-3} - X_{j,2n-1}$ are the soft factors for gluon $n - 1$; as expected, all other tensor elements vanish. 

So, we have now completely determined all terms on the $r.h.s.$ of \cref{eq:NI-form}: all tensors with three or greater indices vanish, and $a_{i,j}$ has only eight non-zero entries --- those listed in \cref{eq:lower-left-indices,eq:upper-right-indices}. Note that, since $a_{i,j}$ must be a skew-symmetric matrix in order to solve \cref{eq:gaug-inv-const}, we only need knowledge of one of \cref{eq:lower-left-indices,eq:upper-right-indices}, because the other is fully determined by the requirement $a_{i,j} = -a_{j,i}$. 

Finally, as we demonstrate in App. D, the double derivative of $\Delta T_2$ in \cref{eq:lower-left-indices} kills all numerators in $\Delta T_2$ except $N_{\mathcal{I}\cup(1,5)}$; see \cref{eq:D-T-simple,eq:D-T-simple-m-3} for the explicit formulae. Of course, the analogous situation holds for the derivatives in \cref{eq:upper-right-indices}, where we find
\begin{equation}
    \frac{\partial^2 \Delta T_{2n-2}}{\partial X_{i,2n} \partial \tilde{\delta}_1} = \frac{\partial^3 N_{\mathcal{I}\cup(2n-1,l)}}{\partial \tilde{\delta}_1 \partial X_{2n-2,l} \partial X_{i,2n}},
\end{equation}
with $l = k$ if $m = 2n-3$ and $l = 2n-5$ otherwise.

\section{Appendix C: Linearity proofs for the $N_\mathcal{I}$}

We will begin by proving that $N_0$ is a linear form in the index $1$. To do this, we use gauge invariance in the $n^{\rm{th}}$ gluon to write
\begin{equation}
    N_0 = \sum_{i = 2}^{2n-2} (X_{i,2n} - X_{i,2n-1}) \frac{\partial N_0}{\partial X_{i,2n}} - \sum_{i} \frac{\partial N_{(i,2n-1)}}{\partial X_{i,2n}}.
\end{equation}
Since $N_0$ is $O(X^2)$, and $i \neq 1$, the first term in the above equation must be linear in $1$. For the second term, we consider each numerator individually, cutting on $X_{i,2n-1} = 0$ to obtain
\begin{equation}
\begin{aligned}
\label{eq:N0-lin-1}
    \frac{\partial N_{(i,2n-1)}}{\partial X_{i,2n}} = \sum_{j\in L,J \in R} (X_{j,J} - X_{j,2n-1} - X_{J,i}) &\frac{\partial^2 N_0^{(L)}}{\partial X_{x_L,j} \partial X_{i,2n}} \frac{\partial N_0^{(R)}}{\partial X_{x_R,J}} - \left( \sum_j \frac{\partial^2 N_0^{(L)}}{\partial X_{x_L,j} \partial X_{i,2n}} \right) \left(\sum_J \frac{\partial N_{(J,i)}^{(R)}}{\partial X_{x_R,J}} \right) \\
    &- \left( \sum_j \frac{\partial^2 N_{(j,2n-1)}^{(L)}}{\partial X_{x_L,j} \partial X_{i,2n}} \right) \left( \sum_J \frac{\partial N_0^{(R)}}{\partial X_{x_R,J}} \right),
\end{aligned}
\end{equation}
with $L = \{ 2n, 1, 2, \ldots, i - 1\}$ and $R = \{ i + 1, i + 2, \ldots, 2n-2\}$. Since the subsurface $R$ does not contain $1$, it is clear by units that the $r.h.s.$ of the above equation must be at most linear in $1$. Hence, $N_0$ is at most linear in $1$.

Now, let us consider sets $\mathcal{I}$ satisfying our original criteria, except now we allow $m = 2n-1$. We want to show that, in this case, $N_\mathcal{I}$ is linear in $1$. We will pursue a inductive strategy, first verifying that, at $n = 4$ points, $N_{(3,7)}$ is linear in $1$. Now, cutting on $X_{k,m} = 0$, we find
\begin{equation}
\label{eq:N-I-fact}
    N_\mathcal{I} = \sum_{j\in D,J \in U} (X_{j,J} - X_{j,m} - X_{J,k}) \frac{\partial N_0^{(D)}}{\partial X_{x_D, j}} \frac{\partial N_{\mathcal{I}_U}^{(U)}}{\partial X_{x_U,J}} -\left( \sum_j \frac{\partial N_0^{(D)}}{\partial X_{x_D, j}} \right) \left( \sum_J \frac{\partial N_{\mathcal{I}_U\cup(J,k)}^{(U)}}{\partial X_{x_U,J}} \right) - \left( \sum_j \frac{\partial N_{(j,m)}^{(D)}}{\partial X_{x_D, j}} \right) \left( \sum_J \frac{\partial N_{\mathcal{I}_U}^{(U)}}{\partial X_{x_U,J}} \right),
\end{equation}
with $U = \{ k+ 1, k+2, \ldots, m - 1\}$, $D \in \{ m + 1, m + 2, \ldots, k - 1\}$, and $\mathcal{I}_U = \mathcal{I} / (k,m)$. In the first line, $1$ appears linearly through $\partial_{X_{x_D,j}}N_0^{(D)}$ when $j \neq 1$. When $j = 1$, it appears linearly in the prefactor, but does not appear in $\partial_{X_{x_D,1}}N_0^{(D)}$, since $N_0^{(D)}$ is linear in $1$. The second line is obviously linear in $1$, through the first term. Finally, in the third line, each term in the $j$-sum for $j \neq 1$ is linear in $1$ by our inductive hypothesis. For $j = 1$, however, we must be a bit more careful. Factorization tells us that
\begin{equation}
\begin{aligned}
\label{eq:j-1-case}
    \frac{\partial N_{(1,m)}^{(D)}}{\partial X_{x_D, 1}} = \sum_{j\in L,J \in R} (X_{j,J} - X_{j,1} - X_{J,m}) &\frac{\partial^2 N_0^{(L)}}{\partial X_{x_D,1} \partial X_{x_L, j}} \frac{\partial N_0^{(R)}}{\partial X_{x_R,J}} - \left( \sum_j  \frac{\partial^2 N_0^{(L)}}{\partial X_{x_D,1} \partial X_{x_L, j}} \right) \left( \sum_J \frac{\partial N_{(J,m)}^{(R)}}{\partial X_{x_R,J}} \right) \\ 
    &- \left( \sum_j \frac{\partial^2 N_{(j,1)}^{(L)}}{\partial X_{x_D,1} \partial X_{x_L, j}} \right) \left( \sum_J \frac{\partial N_0^{(R)}}{\partial X_{x_R,J}} \right),
\end{aligned}
\end{equation}
with $L = \{ 2, 3, \ldots, k, x_D\} $ and $R = \{ m + 1, m + 2, \ldots, 2n\}$. In the first line, $1$ can only appear through terms
\begin{equation}
    X_{j,1} \frac{\partial^2 N_0^{(L)}}{\partial X_{x_D,1} \partial X_{x_L, j}} \left( \sum_J \frac{\partial N_0^{(R)}}{\partial X_{x_R,J}} \right).
\end{equation}
As we discussed in App. A, it was demonstrated in Ref.~\cite{Backus:2025njt} that the $J$-sum in the above is independent of $1$; thus, the term above is at most linear in $1$. In the second term of the first line, the first term is a pure number by units, and, by our inductive hypothesis, the second term is at most linear in $1$. And, finally, the first term in the third line is at most linear in $1$ by units, and the second term is independent of $1$. Thus, \cref{eq:j-1-case} is at most linear in $1$, which finishes the proof that $N_\mathcal{I}$ is independent of $1$ as long as $\mathcal{I}$ does not touch $1$. Of course, the same linearity result holds for any odd index.

Now, let us switch gears and prove that $\partial_{X_{i,2n}} \partial_{X_{2,j}} N_{\mathcal{I}}$ is independent of $1$ and $2n-1$ for all $i = 4, 5, \ldots, 2n-2$ and $j = 4, 5, \ldots, 2n$.

When $\mathcal{I} = 0$, this conclusion is obvious, since $N_0$ is $O(X^2)$. Otherwise, we can cut on $X_{k,m} = 0$, as shown in \cref{eq:N-I-fact}. Then, there are four cases to work through. If $i \in U, j \in U$, $\partial_{X_{i,2n}} \partial_{X_{2,j}} N_{\mathcal{I}}$ vanishes. If $i \in U, j \in D$, then we have
\begin{equation}
    \frac{\partial^2 N_{\mathcal{I}} }{\partial X_{i,2n} \partial X_{2,j}} = \left( \frac{\partial^2 N_0^{(D)}}{\partial X_{x_D, 2n} \partial X_{2,j}} \right) \left( \frac{\partial N_{\mathcal{I}_U}^{(U)}}{\partial X_{x_U,i}} \right).
\end{equation}
This is independent of $1$ and $2n-1$, since all dependence lives in $N_0^{(D)}$, and by units the double derivative reduces this to a pure number. If $i \in D, j \in U$, we find
\begin{equation}
     \frac{\partial^2 N_{\mathcal{I}} }{\partial X_{i,2n} \partial X_{2,j}} = \left( \frac{\partial^2 N_0^{(D)}}{\partial X_{x_D, 2} \partial X_{i,2n}} \right) \left( \frac{\partial^2 N_{\mathcal{I}_U}^{(U)}}{\partial X_{x_U, j}} \right),
\end{equation}
which is independent of $1$ and $2n-1$ for the same reason.

Finally, when $i \in D, j \in D$, factorization tells us that
\begin{equation}
    \frac{\partial^2 N_\mathcal{I}}{\partial X_{i,2n} \partial X_{2, j}} = -\left( \sum_p \frac{\partial N_{\mathcal{I}_U}^{(U)}}{\partial X_{x_U,p}}\right) \left( \sum_P \frac{\partial^3 N_{(P,m)}^{(D)}}{\partial X_{x_D, P} \partial X_{i,2n} \partial X_{2,j}} \right)
\end{equation}
Again by units, this is independent of $2n-1$ and $1$.

Note that, if we allowed $\mathcal{I}$ to touch $2n-1$, $\partial_{X_{i,2n}} \partial_{X_{2,j}} N_{\mathcal{I}}$ would generically depend on $2n-1$ but still be independent of $1$.

Finally, with the above result in-hand, it follows immediately from gauge invariance in gluon $1$ ($e.g.$, \cref{eq:gaug-inv-1,eq:T-3-2-def}) that $\partial_{X_{J,2n}} N_\mathcal{I}$ is independent of $1$ for $J \neq 2,3$. When $J = 2,3$, $\partial_{X_{J,2n}} N_\mathcal{I}$ can be at most linear in $1$. This completes our proofs.

\section{Appendix D: Simplifying the non-zero $a$ matrix elements}

Here we demonstrate that
\begin{equation}
\label{eq:D-T-simple}
    \frac{\partial^2 \Delta T_2}{\partial X_{i,2n} \partial X_{3,2n-1}} = -\frac{\partial^3 N_{\mathcal{I}\cup(1,5)}}{\partial X_{2,5} \partial X_{i,2n} \partial X_{3,2n-1}},
\end{equation}
for $i = 2n-3,2n-2$. Let us begin by considering a set $\mathcal{I}$ which has $k \neq 3$. Since $\Delta T_2 = T_3^{(2)} - T_1^{(2)}$, let us begin with $T_3^{(2)}$, which is written out in \cref{eq:T-3-2-def}. Individually evaluating each $N_{\mathcal{I}\cup(3,j)}$ by cutting on $X_{3,j} = 0$ for $j \neq 2n-1$, we find that $\partial_{X_{2,j}} \partial_{X_{i,2n}} \partial_{X_{3,2n-1}} N_{\mathcal{I}\cup(3,j)} = 0$, since all three derivatives act on the surface below the cut on the momentum polygon. When $j = 2n-1$, this conclusion still holds, because obviously $N_{\mathcal{I}\cup(3,2n-1)}$ is independent of $X_{3,2n-1}$. 

Now, for $T_1^{(2)}$, we use the same strategy. Since cutting on $X_{1,j} = 0$ always splits $N_{\mathcal{I}\cup(1,j)}$ into a surface containing $3$ ($L$) and a surface containing $2n-1$ ($R$), we have
\begin{equation}
\label{eq:T-1-2-intermed}
    \frac{\partial^3 N_{\mathcal{I}\cup(1,j)}}{\partial X_{2,j} \partial X_{3,2n-1} \partial X_{i,2n}} = \left( \frac{ \partial^2 N^{(R)}}{\partial X_{x_R, 2n-1} \partial X_{i,2n}} \right) \left( \frac{\partial^2 N^{(L)}}{\partial X_{2,j} \partial X_{x_L, 3}} \right).
\end{equation}
(We suppress subscripts on the above numerators since they are irrelevant to this proof.) As we showed in App. C, when $j \neq 5$, $\partial_{X_{2, j}} N^{(L)}$ cannot depend on $3$, so \cref{eq:T-1-2-intermed} vanishes. When $j = 5$, the subsurface $L$ is a three-particle surface, and we can explicitly verify that $\partial_{X_{x_L,3}} \partial_{X_{2,5}} \mathcal{A}_3^{(L)} = -1$. So, we have proved \cref{eq:D-T-simple} for $k \neq 3$.

Let us now consider $k = 3$, starting again with $T_3^{(2)}$. In this case, generically there will be terms $N_{\mathcal{I}\cup(3,j)}$ included in $T_3^{(2)}$ where $3 < j < m$. For all other $j$, the above argument goes through, whereas for this new range of $j$, we can cut on $X_{3,m} = 0$, yielding
\begin{equation}
    \frac{\partial^3 N_{\mathcal{I}\cup(3,j)}}{\partial X_{2,j} \partial X_{i,2n} \partial X_{3,2n-1}} = \left( \frac{\partial^3 N_0^{(D)}}{\partial X_{x_D,2} \partial X_{i,2n} \partial X_{3,2n-1}}\right) \cdots.
\end{equation}
Of course, this vanishes by units.

For each term in $T_1^{(2)}$, \cref{eq:T-1-2-intermed} is also valid when $k = 3$. In this case, we can use the fact (also from App. C) that $\partial_{X_{x_L,3}} N^{(L)}$ is independent of $j$ as long as $j \neq m$. If $j = m$, then $L$ is again a three-particle surface where, like before, $\partial_{X_{x_L, 3}} \partial_{X_{2,k}} \mathcal{A}_3^{(L)} = -1$. As a result, when $k = 3$, we have
\begin{equation}
\label{eq:D-T-simple-m-3}
    \frac{\partial^2 \Delta T_2}{\partial X_{i,2n} \partial X_{3,2n-1}} = -\frac{\partial^3 N_{\mathcal{I}\cup(1,m)}}{\partial X_{2,m} \partial X_{i,2n} \partial X_{3,2n-1}}.
\end{equation}

\section{Appendix E: \cref{eq:app-D-gen} when $k = 3$}

In this Appendix, we will prove that
\begin{equation}
\label{eq:App-D-conj}
    \frac{\partial^2 N_\mathcal{I}}{\partial X_{i,2n} \partial X_{2,2n-1}} = -\frac{\partial^2 \Delta T_2}{\partial X_{i,2n} \partial X_{3,2n-1}}
\end{equation}
when $k = 3$ and $i = 2n-3,2n-2$.

Let us start with the $l.h.s.$ of \cref{eq:App-D-conj}. Factorization on $X_{3,m} = 0$ yields
\begin{equation}
    \frac{\partial^2 N_\mathcal{I}}{\partial X_{i,2n} \partial X_{2,2n-1}} = -\left( \sum_J \frac{\partial N_{\mathcal{I}_U}^{(U)}}{\partial X_{x_U,J}} \right)\left( \sum_j \frac{\partial^3 N_{(j,m)}^{(D)}}{\partial X_{x_D,j} \partial X_{i,2n} \partial X_{2,2n-1}} \right),
\end{equation}
and further factorization on $X_{j,m} = 0$ shows that all terms in the $j$-sum vanish except that for $j = 1$. Explicitly, in this case we find
\begin{equation}
\label{eq:first-lhs}
    \frac{\partial^3 N^{(D)}_{(1,m)}}{\partial X_{x_D,1} \partial X_{i,2n} \partial X_{2,2n-1}} = \left( \frac{\partial^2 \mathcal{A}_3^{(L)}}{\partial X_{x_L,2} \partial X_{x_D,1}} \right) \left( \frac{\partial^2 N_0^{(R)}}{\partial X_{x_R,2n-1}\partial X_{i,2n}} \right),
\end{equation}
and the three-point amplitude evaluates to $\partial_{X_{x_D,1}} \partial_{X_{x_L,2}} \mathcal{A}_3^{(L)} = 1$. 

We proved in App. D that the $r.h.s.$ of \cref{eq:App-D-conj} is equal to $\partial_{X_{2,m}} \partial_{X_{3,2n-1}} \partial_{X_{i,2n}} N_{\mathcal{I}\cup(1,m)}$. Cutting on $X_{3,m} = 0$ yields
\begin{equation}
\label{eq:first-rhs}
    \frac{\partial^3 N_{\mathcal{I}\cup(1,m)}}{\partial X_{2,m} \partial X_{i,2n} \partial X_{3,2n-1}} = -\left( \sum_J \frac{\partial N_{\mathcal{I}_U}^{(U)}}{\partial X_{x_U, J}} \right) \left( \frac{\partial^3 N_{(1,m)}^{(D)}}{\partial X_{x_D,2n-1} \partial X_{i,2n} \partial X_{2,m}} \right),
\end{equation}
and further cutting on $X_{1,m} = 0$ gives
\begin{equation}
\label{eq:last-rhs}
    \frac{\partial^3 N_{(1,m)}^{(D)}}{\partial X_{x_D,2n-1} \partial X_{i,2n} \partial X_{2,m}} = \left( \frac{\partial^2 \mathcal{A}_3^{(L)}}{\partial X_{x_L,x_D} \partial X_{2,m}} \right) \left( \frac{\partial^2 N_0^{(R)}}{\partial X_{x_R,2n-1}\partial X_{i,2n}} \right).
\end{equation}
Now, using the three-particle amplitude, we find $\partial_{X_{x_L, x_D}} \partial_{X_{2,m}} \mathcal{A}_3^{(L)} = 1$. Thus, \cref{eq:last-rhs} matches \cref{eq:first-lhs}, which means that $\partial_{X_{i,2n}} \partial_{X_{2,2n-1}} N_\mathcal{I}$ is equal to \cref{eq:first-rhs}. This completes the proof.

\section{Appendix F: Contact team recursion relations}

Here we give recursive formulae that determine the contact term of the $n$-point amplitude purely from contact terms of lower-point amplitudes. These follow as a direct consequence of \cref{eq:final-recur-1,eq:final-recur-2} and factorization, though we do not supply a proof now.

As we know from \cref{eq:N-gauge-inv}, the $n$-point contact term $N^{(n)}_0(1,2,\ldots,2n)$ may be written in the form 
\begin{equation}
\label{eq:contact-form}
    N^{(n)}_0 = \sum_{j = 2}^{2n - 2} (X_{j,2n} - X_{j,2n-1})C_j - T_{2n-1}^{(2n)}, 
\end{equation}
with
\begin{equation}
    T_{2n-1}^{(2n)} = \sum_{j = 3, 5, \ldots, 2n-5} \frac{\partial N^{(n)}_{(j,2n-1)}}{\partial X_{j,2n}}, \quad C_j = \frac{\partial N^{(n)}_0}{\partial X_{j,2n}}. 
\end{equation}
Using factorization, one can show that
\begin{equation}
\begin{aligned}
\label{eq:contact-T}
    T_{2n-1}^{(2n)} &= -N^{(n-1)'}_0 + \sum_{j = 4}^{2n-2} (\delta_2 - \delta_j)\frac{\partial N^{(n-1)'}_0}{\partial X_{2,j}} + (\delta_2 - \delta_3) \sum_{k \in R_1} \mathcal{W}_x[N_0(k,k+1,\ldots,2n-1,x)] \\
    -& \sum_{k \in R_1, i \in R_2} (\mathcal{W}_x[N_0(k,k+1,\ldots,2n-1,x)])\cdot(\mathcal{W}_y[N_0(1,2,\ldots,i,y)]), 
\end{aligned}
\end{equation}
where $N^{(n-1)'}_0 = N_0(2n-1,2,\ldots,2n-2)$, $R_1 = \{ 5, 7, \ldots, 2n-5\}$, and $R_2 = \{ 5, 7, \ldots, k - 2\}$. Additionally, we have
\begin{equation}
    \mathcal{W}_{2n}[N_0(1,2,\ldots,2n)] = \sum_{j = 2}^{2n} \frac{\partial N_0}{\partial X_{j,2n}},
\end{equation}
an operator which was introduced in Ref.~\cite{Backus:2025njt}.

With these results in-hand, gauge invariance dictates the form of the $C_j$:
\begin{equation}
\begin{aligned}
\label{eq:cont-term-Cs}
    &C_2 = \frac{\partial N^{(n-1)}_0}{\partial X_{2,2n-2}} - \sum_{j = 4}^{2n-2} \frac{\partial N^{(n-1)'}_0}{\partial X_{2,j}} - \sum_{k \in R_1} \mathcal{W}_x[N_0(k,k+1,\ldots,2n-1,x)] + \delta_{2n-3} - \delta_{2n-2}, \\
    &C_3 = \frac{\partial N^{(n-1)}_0}{\partial X_{3,2n-2}} + \sum_{k \in R_1} \mathcal{W}_x[N_0(k,k+1,\ldots,2n-1,x)] - \delta_{2n-3} + \delta_{2n-2}, \\
    &C_i = \frac{\partial N_0^{(n-1)'}}{\partial X_{i,2n-1}} + \frac{\partial N_0^{(n-1)}}{\partial X_{i,2n-2}} + \frac{\partial N_0^{(n-1)'}}{\partial X_{i,2}}, \quad i = 4, 5, \ldots, 2n-4, \\
    &C_{2n-3} = \frac{\partial N^{(n-1)'}_0}{\partial X_{2,2n-3}} + \sum_{k \in R_1} \mathcal{W}_y[N_0(1,2,\ldots,k,y)] + \delta_3 - \delta_2, \\
    &C_{2n-2} = \frac{\partial N^{(n-1)'}_0}{\partial X_{2,2n-2}} - \sum_{j = 2}^{2n-4} \frac{\partial N^{(n-1)}_0}{\partial X_{j,2n-2}} - \sum_{k \in R_1} \mathcal{W}_y[N_0(1,2,\ldots,k,y)] - \delta_3 + \delta_2.
\end{aligned}
\end{equation}
Thus, by plugging \cref{eq:cont-term-Cs,eq:contact-T} into \cref{eq:contact-form}, we derive a purely contact-term recursion, starting from knowledge only of the three-particle amplitude.

\bibliographystyle{apsrev4-1}
\bibliography{ref}{}

\begin{thebibliography}{39}%
\makeatletter
\providecommand \@ifxundefined [1]{%
 \@ifx{#1\undefined}
}%
\providecommand \@ifnum [1]{%
 \ifnum #1\expandafter \@firstoftwo
 \else \expandafter \@secondoftwo
 \fi
}%
\providecommand \@ifx [1]{%
 \ifx #1\expandafter \@firstoftwo
 \else \expandafter \@secondoftwo
 \fi
}%
\providecommand \natexlab [1]{#1}%
\providecommand \enquote  [1]{``#1''}%
\providecommand \bibnamefont  [1]{#1}%
\providecommand \bibfnamefont [1]{#1}%
\providecommand \citenamefont [1]{#1}%
\providecommand \href@noop [0]{\@secondoftwo}%
\providecommand \href [0]{\begingroup \@sanitize@url \@href}%
\providecommand \@href[1]{\@@startlink{#1}\@@href}%
\providecommand \@@href[1]{\endgroup#1\@@endlink}%
\providecommand \@sanitize@url [0]{\catcode `\\12\catcode `\$12\catcode
  `\&12\catcode `\#12\catcode `\^12\catcode `\_12\catcode `\%12\relax}%
\providecommand \@@startlink[1]{}%
\providecommand \@@endlink[0]{}%
\providecommand \url  [0]{\begingroup\@sanitize@url \@url }%
\providecommand \@url [1]{\endgroup\@href {#1}{\urlprefix }}%
\providecommand \urlprefix  [0]{URL }%
\providecommand \Eprint [0]{\href }%
\providecommand \doibase [0]{http://dx.doi.org/}%
\providecommand \selectlanguage [0]{\@gobble}%
\providecommand \bibinfo  [0]{\@secondoftwo}%
\providecommand \bibfield  [0]{\@secondoftwo}%
\providecommand \translation [1]{[#1]}%
\providecommand \BibitemOpen [0]{}%
\providecommand \bibitemStop [0]{}%
\providecommand \bibitemNoStop [0]{.\EOS\space}%
\providecommand \EOS [0]{\spacefactor3000\relax}%
\providecommand \BibitemShut  [1]{\csname bibitem#1\endcsname}%
\let\auto@bib@innerbib\@empty
\bibitem [{\citenamefont {Arkani-Hamed}\ \emph
  {et~al.}(2025{\natexlab{a}})\citenamefont {Arkani-Hamed}, \citenamefont
  {Frost}, \citenamefont {Salvatori}, \citenamefont {Plamondon},\ and\
  \citenamefont {Thomas}}]{Arkani-Hamed:2023lbd}%
  \BibitemOpen
  \bibfield  {author} {\bibinfo {author} {\bibfnamefont {N.}~\bibnamefont
  {Arkani-Hamed}}, \bibinfo {author} {\bibfnamefont {H.}~\bibnamefont {Frost}},
  \bibinfo {author} {\bibfnamefont {G.}~\bibnamefont {Salvatori}}, \bibinfo
  {author} {\bibfnamefont {P.-G.}\ \bibnamefont {Plamondon}}, \ and\ \bibinfo
  {author} {\bibfnamefont {H.}~\bibnamefont {Thomas}},\ }\href {\doibase
  10.1007/JHEP08(2025)194} {\bibfield  {journal} {\bibinfo  {journal} {JHEP}\
  }\textbf {\bibinfo {volume} {08}},\ \bibinfo {pages} {194} (\bibinfo {year}
  {2025}{\natexlab{a}})},\ \Eprint {http://arxiv.org/abs/2309.15913}
  {arXiv:2309.15913 [hep-th]} \BibitemShut {NoStop}%
\bibitem [{\citenamefont {Arkani-Hamed}\ \emph
  {et~al.}(2025{\natexlab{b}})\citenamefont {Arkani-Hamed}, \citenamefont
  {Frost}, \citenamefont {Salvatori}, \citenamefont {Plamondon},\ and\
  \citenamefont {Thomas}}]{Arkani-Hamed:2023mvg}%
  \BibitemOpen
  \bibfield  {author} {\bibinfo {author} {\bibfnamefont {N.}~\bibnamefont
  {Arkani-Hamed}}, \bibinfo {author} {\bibfnamefont {H.}~\bibnamefont {Frost}},
  \bibinfo {author} {\bibfnamefont {G.}~\bibnamefont {Salvatori}}, \bibinfo
  {author} {\bibfnamefont {P.-G.}\ \bibnamefont {Plamondon}}, \ and\ \bibinfo
  {author} {\bibfnamefont {H.}~\bibnamefont {Thomas}},\ }\href {\doibase
  10.1007/JHEP09(2025)033} {\bibfield  {journal} {\bibinfo  {journal} {JHEP}\
  }\textbf {\bibinfo {volume} {09}},\ \bibinfo {pages} {033} (\bibinfo {year}
  {2025}{\natexlab{b}})},\ \Eprint {http://arxiv.org/abs/2311.09284}
  {arXiv:2311.09284 [hep-th]} \BibitemShut {NoStop}%
\bibitem [{\citenamefont {Arkani-Hamed}\ \emph
  {et~al.}(2024{\natexlab{a}})\citenamefont {Arkani-Hamed}, \citenamefont
  {Cao}, \citenamefont {Dong}, \citenamefont {Figueiredo},\ and\ \citenamefont
  {He}}]{Arkani-Hamed:2023swr}%
  \BibitemOpen
  \bibfield  {author} {\bibinfo {author} {\bibfnamefont {N.}~\bibnamefont
  {Arkani-Hamed}}, \bibinfo {author} {\bibfnamefont {Q.}~\bibnamefont {Cao}},
  \bibinfo {author} {\bibfnamefont {J.}~\bibnamefont {Dong}}, \bibinfo {author}
  {\bibfnamefont {C.}~\bibnamefont {Figueiredo}}, \ and\ \bibinfo {author}
  {\bibfnamefont {S.}~\bibnamefont {He}},\ }\href {\doibase
  10.1007/JHEP10(2024)231} {\bibfield  {journal} {\bibinfo  {journal} {JHEP}\
  }\textbf {\bibinfo {volume} {10}},\ \bibinfo {pages} {231} (\bibinfo {year}
  {2024}{\natexlab{a}})},\ \Eprint {http://arxiv.org/abs/2312.16282}
  {arXiv:2312.16282 [hep-th]} \BibitemShut {NoStop}%
\bibitem [{\citenamefont {Arkani-Hamed}\ \emph
  {et~al.}(2025{\natexlab{c}})\citenamefont {Arkani-Hamed}, \citenamefont
  {Cao}, \citenamefont {Dong}, \citenamefont {Figueiredo},\ and\ \citenamefont
  {He}}]{Arkani-Hamed:2023jry}%
  \BibitemOpen
  \bibfield  {author} {\bibinfo {author} {\bibfnamefont {N.}~\bibnamefont
  {Arkani-Hamed}}, \bibinfo {author} {\bibfnamefont {Q.}~\bibnamefont {Cao}},
  \bibinfo {author} {\bibfnamefont {J.}~\bibnamefont {Dong}}, \bibinfo {author}
  {\bibfnamefont {C.}~\bibnamefont {Figueiredo}}, \ and\ \bibinfo {author}
  {\bibfnamefont {S.}~\bibnamefont {He}},\ }\href {\doibase
  10.1007/JHEP04(2025)078} {\bibfield  {journal} {\bibinfo  {journal} {JHEP}\
  }\textbf {\bibinfo {volume} {04}},\ \bibinfo {pages} {078} (\bibinfo {year}
  {2025}{\natexlab{c}})},\ \Eprint {http://arxiv.org/abs/2401.00041}
  {arXiv:2401.00041 [hep-th]} \BibitemShut {NoStop}%
\bibitem [{\citenamefont {Arkani-Hamed}\ \emph
  {et~al.}(2024{\natexlab{b}})\citenamefont {Arkani-Hamed}, \citenamefont
  {Cao}, \citenamefont {Dong}, \citenamefont {Figueiredo},\ and\ \citenamefont
  {He}}]{Arkani-Hamed:2024nhp}%
  \BibitemOpen
  \bibfield  {author} {\bibinfo {author} {\bibfnamefont {N.}~\bibnamefont
  {Arkani-Hamed}}, \bibinfo {author} {\bibfnamefont {Q.}~\bibnamefont {Cao}},
  \bibinfo {author} {\bibfnamefont {J.}~\bibnamefont {Dong}}, \bibinfo {author}
  {\bibfnamefont {C.}~\bibnamefont {Figueiredo}}, \ and\ \bibinfo {author}
  {\bibfnamefont {S.}~\bibnamefont {He}},\ }\href {\doibase
  10.1103/PhysRevD.110.065018} {\bibfield  {journal} {\bibinfo  {journal}
  {Phys. Rev. D}\ }\textbf {\bibinfo {volume} {110}},\ \bibinfo {pages}
  {065018} (\bibinfo {year} {2024}{\natexlab{b}})},\ \Eprint
  {http://arxiv.org/abs/2401.05483} {arXiv:2401.05483 [hep-th]} \BibitemShut
  {NoStop}%
\bibitem [{\citenamefont {Arkani-Hamed}\ \emph
  {et~al.}(2025{\natexlab{d}})\citenamefont {Arkani-Hamed}, \citenamefont
  {Figueiredo}, \citenamefont {Frost},\ and\ \citenamefont
  {Salvatori}}]{Arkani-Hamed:2024vna}%
  \BibitemOpen
  \bibfield  {author} {\bibinfo {author} {\bibfnamefont {N.}~\bibnamefont
  {Arkani-Hamed}}, \bibinfo {author} {\bibfnamefont {C.}~\bibnamefont
  {Figueiredo}}, \bibinfo {author} {\bibfnamefont {H.}~\bibnamefont {Frost}}, \
  and\ \bibinfo {author} {\bibfnamefont {G.}~\bibnamefont {Salvatori}},\ }\href
  {\doibase 10.1007/JHEP05(2025)051} {\bibfield  {journal} {\bibinfo  {journal}
  {JHEP}\ }\textbf {\bibinfo {volume} {05}},\ \bibinfo {pages} {051} (\bibinfo
  {year} {2025}{\natexlab{d}})},\ \Eprint {http://arxiv.org/abs/2402.06719}
  {arXiv:2402.06719 [hep-th]} \BibitemShut {NoStop}%
\bibitem [{\citenamefont {Arkani-Hamed}\ and\ \citenamefont
  {Figueiredo}(2025{\natexlab{a}})}]{Arkani-Hamed:2024yvu}%
  \BibitemOpen
  \bibfield  {author} {\bibinfo {author} {\bibfnamefont {N.}~\bibnamefont
  {Arkani-Hamed}}\ and\ \bibinfo {author} {\bibfnamefont {C.}~\bibnamefont
  {Figueiredo}},\ }\href {\doibase 10.1007/JHEP09(2025)189} {\bibfield
  {journal} {\bibinfo  {journal} {JHEP}\ }\textbf {\bibinfo {volume} {09}},\
  \bibinfo {pages} {189} (\bibinfo {year} {2025}{\natexlab{a}})},\ \Eprint
  {http://arxiv.org/abs/2403.04826} {arXiv:2403.04826 [hep-th]} \BibitemShut
  {NoStop}%
\bibitem [{\citenamefont {Arkani-Hamed}\ and\ \citenamefont
  {Figueiredo}(2025{\natexlab{b}})}]{Arkani-Hamed:2024fyd}%
  \BibitemOpen
  \bibfield  {author} {\bibinfo {author} {\bibfnamefont {N.}~\bibnamefont
  {Arkani-Hamed}}\ and\ \bibinfo {author} {\bibfnamefont {C.}~\bibnamefont
  {Figueiredo}},\ }\href {\doibase 10.1007/JHEP10(2025)077} {\bibfield
  {journal} {\bibinfo  {journal} {JHEP}\ }\textbf {\bibinfo {volume} {10}},\
  \bibinfo {pages} {077} (\bibinfo {year} {2025}{\natexlab{b}})},\ \Eprint
  {http://arxiv.org/abs/2405.09608} {arXiv:2405.09608 [hep-th]} \BibitemShut
  {NoStop}%
\bibitem [{\citenamefont {Arkani-Hamed}\ \emph
  {et~al.}(2025{\natexlab{e}})\citenamefont {Arkani-Hamed}, \citenamefont
  {Cao}, \citenamefont {Dong}, \citenamefont {Figueiredo},\ and\ \citenamefont
  {He}}]{Arkani-Hamed:2024tzl}%
  \BibitemOpen
  \bibfield  {author} {\bibinfo {author} {\bibfnamefont {N.}~\bibnamefont
  {Arkani-Hamed}}, \bibinfo {author} {\bibfnamefont {Q.}~\bibnamefont {Cao}},
  \bibinfo {author} {\bibfnamefont {J.}~\bibnamefont {Dong}}, \bibinfo {author}
  {\bibfnamefont {C.}~\bibnamefont {Figueiredo}}, \ and\ \bibinfo {author}
  {\bibfnamefont {S.}~\bibnamefont {He}},\ }\href {\doibase
  10.1103/PhysRevLett.134.171601} {\bibfield  {journal} {\bibinfo  {journal}
  {Phys. Rev. Lett.}\ }\textbf {\bibinfo {volume} {134}},\ \bibinfo {pages}
  {171601} (\bibinfo {year} {2025}{\natexlab{e}})},\ \Eprint
  {http://arxiv.org/abs/2408.11891} {arXiv:2408.11891 [hep-th]} \BibitemShut
  {NoStop}%
\bibitem [{\citenamefont {Arkani-Hamed}\ \emph
  {et~al.}(2024{\natexlab{c}})\citenamefont {Arkani-Hamed}, \citenamefont
  {Frost},\ and\ \citenamefont {Salvatori}}]{Arkani-Hamed:2024pzc}%
  \BibitemOpen
  \bibfield  {author} {\bibinfo {author} {\bibfnamefont {N.}~\bibnamefont
  {Arkani-Hamed}}, \bibinfo {author} {\bibfnamefont {H.}~\bibnamefont {Frost}},
  \ and\ \bibinfo {author} {\bibfnamefont {G.}~\bibnamefont {Salvatori}},\
  }\href@noop {} {\  (\bibinfo {year} {2024}{\natexlab{c}})},\ \Eprint
  {http://arxiv.org/abs/2412.21027} {arXiv:2412.21027 [hep-th]} \BibitemShut
  {NoStop}%
\bibitem [{\citenamefont {Backus}\ and\ \citenamefont
  {Figueiredo}(2025)}]{Backus:2025njt}%
  \BibitemOpen
  \bibfield  {author} {\bibinfo {author} {\bibfnamefont {J.~V.}\ \bibnamefont
  {Backus}}\ and\ \bibinfo {author} {\bibfnamefont {C.}~\bibnamefont
  {Figueiredo}},\ }\href {\doibase 10.1007/JHEP09(2025)069} {\bibfield
  {journal} {\bibinfo  {journal} {JHEP}\ }\textbf {\bibinfo {volume} {09}},\
  \bibinfo {pages} {069} (\bibinfo {year} {2025})},\ \Eprint
  {http://arxiv.org/abs/2505.17179} {arXiv:2505.17179 [hep-th]} \BibitemShut
  {NoStop}%
\bibitem [{\citenamefont {Cao}\ \emph {et~al.}(2025)\citenamefont {Cao},
  \citenamefont {Dong}, \citenamefont {He},\ and\ \citenamefont
  {Zhu}}]{Cao:2025lzv}%
  \BibitemOpen
  \bibfield  {author} {\bibinfo {author} {\bibfnamefont {Q.}~\bibnamefont
  {Cao}}, \bibinfo {author} {\bibfnamefont {J.}~\bibnamefont {Dong}}, \bibinfo
  {author} {\bibfnamefont {S.}~\bibnamefont {He}}, \ and\ \bibinfo {author}
  {\bibfnamefont {F.}~\bibnamefont {Zhu}},\ }\href@noop {} {\  (\bibinfo {year}
  {2025})},\ \Eprint {http://arxiv.org/abs/2504.21676} {arXiv:2504.21676
  [hep-th]} \BibitemShut {NoStop}%
\bibitem [{\citenamefont {De}\ \emph {et~al.}(2024)\citenamefont {De},
  \citenamefont {Pokraka}, \citenamefont {Skowronek}, \citenamefont
  {Spradlin},\ and\ \citenamefont {Volovich}}]{De:2024wsy}%
  \BibitemOpen
  \bibfield  {author} {\bibinfo {author} {\bibfnamefont {S.}~\bibnamefont
  {De}}, \bibinfo {author} {\bibfnamefont {A.}~\bibnamefont {Pokraka}},
  \bibinfo {author} {\bibfnamefont {M.}~\bibnamefont {Skowronek}}, \bibinfo
  {author} {\bibfnamefont {M.}~\bibnamefont {Spradlin}}, \ and\ \bibinfo
  {author} {\bibfnamefont {A.}~\bibnamefont {Volovich}},\ }\href {\doibase
  10.1007/JHEP09(2024)160} {\bibfield  {journal} {\bibinfo  {journal} {JHEP}\
  }\textbf {\bibinfo {volume} {09}},\ \bibinfo {pages} {160} (\bibinfo {year}
  {2024})},\ \Eprint {http://arxiv.org/abs/2406.04411} {arXiv:2406.04411
  [hep-th]} \BibitemShut {NoStop}%
\bibitem [{\citenamefont {Salvatori}(2025)}]{Salvatori:2025oib}%
  \BibitemOpen
  \bibfield  {author} {\bibinfo {author} {\bibfnamefont {G.}~\bibnamefont
  {Salvatori}},\ }\href@noop {} {\  (\bibinfo {year} {2025})},\ \Eprint
  {http://arxiv.org/abs/2503.07707} {arXiv:2503.07707 [hep-th]} \BibitemShut
  {NoStop}%
\bibitem [{\citenamefont {Arkani-Hamed}\ \emph
  {et~al.}(2018{\natexlab{a}})\citenamefont {Arkani-Hamed}, \citenamefont
  {Bai}, \citenamefont {He},\ and\ \citenamefont {Yan}}]{Arkani-Hamed:2017mur}%
  \BibitemOpen
  \bibfield  {author} {\bibinfo {author} {\bibfnamefont {N.}~\bibnamefont
  {Arkani-Hamed}}, \bibinfo {author} {\bibfnamefont {Y.}~\bibnamefont {Bai}},
  \bibinfo {author} {\bibfnamefont {S.}~\bibnamefont {He}}, \ and\ \bibinfo
  {author} {\bibfnamefont {G.}~\bibnamefont {Yan}},\ }\href {\doibase
  10.1007/JHEP05(2018)096} {\bibfield  {journal} {\bibinfo  {journal} {JHEP}\
  }\textbf {\bibinfo {volume} {05}},\ \bibinfo {pages} {096} (\bibinfo {year}
  {2018}{\natexlab{a}})},\ \Eprint {http://arxiv.org/abs/1711.09102}
  {arXiv:1711.09102 [hep-th]} \BibitemShut {NoStop}%
\bibitem [{\citenamefont {Arkani-Hamed}\ \emph {et~al.}(2022)\citenamefont
  {Arkani-Hamed}, \citenamefont {He}, \citenamefont {Salvatori},\ and\
  \citenamefont {Thomas}}]{Arkani-Hamed:2019vag}%
  \BibitemOpen
  \bibfield  {author} {\bibinfo {author} {\bibfnamefont {N.}~\bibnamefont
  {Arkani-Hamed}}, \bibinfo {author} {\bibfnamefont {S.}~\bibnamefont {He}},
  \bibinfo {author} {\bibfnamefont {G.}~\bibnamefont {Salvatori}}, \ and\
  \bibinfo {author} {\bibfnamefont {H.}~\bibnamefont {Thomas}},\ }\href
  {\doibase 10.1007/JHEP11(2022)049} {\bibfield  {journal} {\bibinfo  {journal}
  {JHEP}\ }\textbf {\bibinfo {volume} {11}},\ \bibinfo {pages} {049} (\bibinfo
  {year} {2022})},\ \Eprint {http://arxiv.org/abs/1912.12948} {arXiv:1912.12948
  [hep-th]} \BibitemShut {NoStop}%
\bibitem [{\citenamefont {Rodina}(2025)}]{Rodina:2024yfc}%
  \BibitemOpen
  \bibfield  {author} {\bibinfo {author} {\bibfnamefont {L.}~\bibnamefont
  {Rodina}},\ }\href {\doibase 10.1103/PhysRevLett.134.031601} {\bibfield
  {journal} {\bibinfo  {journal} {Phys. Rev. Lett.}\ }\textbf {\bibinfo
  {volume} {134}},\ \bibinfo {pages} {031601} (\bibinfo {year} {2025})},\
  \Eprint {http://arxiv.org/abs/2406.04234} {arXiv:2406.04234 [hep-th]}
  \BibitemShut {NoStop}%
\bibitem [{\citenamefont {Backus}\ and\ \citenamefont
  {Rodina}(2025)}]{Backus:2025hpn}%
  \BibitemOpen
  \bibfield  {author} {\bibinfo {author} {\bibfnamefont {J.~V.}\ \bibnamefont
  {Backus}}\ and\ \bibinfo {author} {\bibfnamefont {L.}~\bibnamefont
  {Rodina}},\ }\href {\doibase 10.1103/7lwk-m5yb} {\bibfield  {journal}
  {\bibinfo  {journal} {Phys. Rev. Lett.}\ }\textbf {\bibinfo {volume} {135}},\
  \bibinfo {pages} {131601} (\bibinfo {year} {2025})},\ \Eprint
  {http://arxiv.org/abs/2503.03805} {arXiv:2503.03805 [hep-th]} \BibitemShut
  {NoStop}%
\bibitem [{\citenamefont {Jones}\ and\ \citenamefont
  {Paranjape}(2025)}]{Jones:2025rbv}%
  \BibitemOpen
  \bibfield  {author} {\bibinfo {author} {\bibfnamefont {C.~R.~T.}\
  \bibnamefont {Jones}}\ and\ \bibinfo {author} {\bibfnamefont
  {S.}~\bibnamefont {Paranjape}},\ }\href {\doibase 10.1007/JHEP07(2025)251}
  {\bibfield  {journal} {\bibinfo  {journal} {JHEP}\ }\textbf {\bibinfo
  {volume} {07}},\ \bibinfo {pages} {251} (\bibinfo {year} {2025})},\ \Eprint
  {http://arxiv.org/abs/2505.02520} {arXiv:2505.02520 [hep-th]} \BibitemShut
  {NoStop}%
\bibitem [{\citenamefont {Paranjape}\ \emph {et~al.}(2025)\citenamefont
  {Paranjape}, \citenamefont {Skowronek}, \citenamefont {Spradlin},\ and\
  \citenamefont {Volovich}}]{Paranjape:2025wjk}%
  \BibitemOpen
  \bibfield  {author} {\bibinfo {author} {\bibfnamefont {S.}~\bibnamefont
  {Paranjape}}, \bibinfo {author} {\bibfnamefont {M.}~\bibnamefont
  {Skowronek}}, \bibinfo {author} {\bibfnamefont {M.}~\bibnamefont {Spradlin}},
  \ and\ \bibinfo {author} {\bibfnamefont {A.}~\bibnamefont {Volovich}},\
  }\href@noop {} {\  (\bibinfo {year} {2025})},\ \Eprint
  {http://arxiv.org/abs/2504.11253} {arXiv:2504.11253 [hep-th]} \BibitemShut
  {NoStop}%
\bibitem [{\citenamefont {Feng}\ \emph {et~al.}(2025)\citenamefont {Feng},
  \citenamefont {Zhang},\ and\ \citenamefont {Zhou}}]{Feng:2025dci}%
  \BibitemOpen
  \bibfield  {author} {\bibinfo {author} {\bibfnamefont {B.}~\bibnamefont
  {Feng}}, \bibinfo {author} {\bibfnamefont {L.}~\bibnamefont {Zhang}}, \ and\
  \bibinfo {author} {\bibfnamefont {K.}~\bibnamefont {Zhou}},\ }\href@noop {}
  {\  (\bibinfo {year} {2025})},\ \Eprint {http://arxiv.org/abs/2508.21345}
  {arXiv:2508.21345 [hep-th]} \BibitemShut {NoStop}%
\bibitem [{\citenamefont {Li}\ and\ \citenamefont {Zhou}(2025)}]{Li:2025suo}%
  \BibitemOpen
  \bibfield  {author} {\bibinfo {author} {\bibfnamefont {X.}~\bibnamefont
  {Li}}\ and\ \bibinfo {author} {\bibfnamefont {K.}~\bibnamefont {Zhou}},\
  }\href@noop {} {\  (\bibinfo {year} {2025})},\ \Eprint
  {http://arxiv.org/abs/2508.12894} {arXiv:2508.12894 [hep-th]} \BibitemShut
  {NoStop}%
\bibitem [{\citenamefont {Zhou}(2025{\natexlab{a}})}]{Zhou:2024ddy}%
  \BibitemOpen
  \bibfield  {author} {\bibinfo {author} {\bibfnamefont {K.}~\bibnamefont
  {Zhou}},\ }\href {\doibase 10.1007/JHEP03(2025)154} {\bibfield  {journal}
  {\bibinfo  {journal} {JHEP}\ }\textbf {\bibinfo {volume} {03}},\ \bibinfo
  {pages} {154} (\bibinfo {year} {2025}{\natexlab{a}})},\ \Eprint
  {http://arxiv.org/abs/2411.07944} {arXiv:2411.07944 [hep-th]} \BibitemShut
  {NoStop}%
\bibitem [{\citenamefont {Bartsch}\ \emph {et~al.}(2025)\citenamefont
  {Bartsch}, \citenamefont {Brown}, \citenamefont {Kampf}, \citenamefont
  {Oktem}, \citenamefont {Paranjape},\ and\ \citenamefont
  {Trnka}}]{Bartsch:2024amu}%
  \BibitemOpen
  \bibfield  {author} {\bibinfo {author} {\bibfnamefont {C.}~\bibnamefont
  {Bartsch}}, \bibinfo {author} {\bibfnamefont {T.~V.}\ \bibnamefont {Brown}},
  \bibinfo {author} {\bibfnamefont {K.}~\bibnamefont {Kampf}}, \bibinfo
  {author} {\bibfnamefont {U.}~\bibnamefont {Oktem}}, \bibinfo {author}
  {\bibfnamefont {S.}~\bibnamefont {Paranjape}}, \ and\ \bibinfo {author}
  {\bibfnamefont {J.}~\bibnamefont {Trnka}},\ }\href {\doibase
  10.1103/PhysRevD.111.045019} {\bibfield  {journal} {\bibinfo  {journal}
  {Phys. Rev. D}\ }\textbf {\bibinfo {volume} {111}},\ \bibinfo {pages}
  {045019} (\bibinfo {year} {2025})},\ \Eprint
  {http://arxiv.org/abs/2403.10594} {arXiv:2403.10594 [hep-th]} \BibitemShut
  {NoStop}%
\bibitem [{\citenamefont {Zhou}(2025{\natexlab{b}})}]{Zhou:2025tvq}%
  \BibitemOpen
  \bibfield  {author} {\bibinfo {author} {\bibfnamefont {K.}~\bibnamefont
  {Zhou}},\ }\href@noop {} {\  (\bibinfo {year} {2025}{\natexlab{b}})},\
  \Eprint {http://arxiv.org/abs/2510.11070} {arXiv:2510.11070 [hep-th]}
  \BibitemShut {NoStop}%
\bibitem [{\citenamefont {Li}\ \emph {et~al.}(2025)\citenamefont {Li},
  \citenamefont {Roest},\ and\ \citenamefont {ter Veldhuis}}]{Li:2024qfp}%
  \BibitemOpen
  \bibfield  {author} {\bibinfo {author} {\bibfnamefont {Y.}~\bibnamefont
  {Li}}, \bibinfo {author} {\bibfnamefont {D.}~\bibnamefont {Roest}}, \ and\
  \bibinfo {author} {\bibfnamefont {T.}~\bibnamefont {ter Veldhuis}},\ }\href
  {\doibase 10.1007/JHEP04(2025)121} {\bibfield  {journal} {\bibinfo  {journal}
  {JHEP}\ }\textbf {\bibinfo {volume} {04}},\ \bibinfo {pages} {121} (\bibinfo
  {year} {2025})},\ \Eprint {http://arxiv.org/abs/2403.12939} {arXiv:2403.12939
  [hep-th]} \BibitemShut {NoStop}%
\bibitem [{\citenamefont {De}\ \emph {et~al.}(2025)\citenamefont {De},
  \citenamefont {Paranjape}, \citenamefont {Pokraka}, \citenamefont
  {Spradlin},\ and\ \citenamefont {Volovich}}]{De:2025bmf}%
  \BibitemOpen
  \bibfield  {author} {\bibinfo {author} {\bibfnamefont {S.}~\bibnamefont
  {De}}, \bibinfo {author} {\bibfnamefont {S.}~\bibnamefont {Paranjape}},
  \bibinfo {author} {\bibfnamefont {A.}~\bibnamefont {Pokraka}}, \bibinfo
  {author} {\bibfnamefont {M.}~\bibnamefont {Spradlin}}, \ and\ \bibinfo
  {author} {\bibfnamefont {A.}~\bibnamefont {Volovich}},\ }\href {\doibase
  10.1007/JHEP07(2025)174} {\bibfield  {journal} {\bibinfo  {journal} {JHEP}\
  }\textbf {\bibinfo {volume} {07}},\ \bibinfo {pages} {174} (\bibinfo {year}
  {2025})},\ \Eprint {http://arxiv.org/abs/2503.23579} {arXiv:2503.23579
  [hep-th]} \BibitemShut {NoStop}%
\bibitem [{\citenamefont {Figueiredo}\ and\ \citenamefont
  {Vaz{\~a}o}(2025)}]{Figueiredo:2025daa}%
  \BibitemOpen
  \bibfield  {author} {\bibinfo {author} {\bibfnamefont {C.}~\bibnamefont
  {Figueiredo}}\ and\ \bibinfo {author} {\bibfnamefont {F.}~\bibnamefont
  {Vaz{\~a}o}},\ }\href@noop {} {\  (\bibinfo {year} {2025})},\ \Eprint
  {http://arxiv.org/abs/2506.19907} {arXiv:2506.19907 [hep-th]} \BibitemShut
  {NoStop}%
\bibitem [{\citenamefont {Arkani-Hamed}\ \emph
  {et~al.}(2024{\natexlab{d}})\citenamefont {Arkani-Hamed}, \citenamefont
  {Figueiredo},\ and\ \citenamefont {Vaz{\~a}o}}]{Arkani-Hamed:2024jbp}%
  \BibitemOpen
  \bibfield  {author} {\bibinfo {author} {\bibfnamefont {N.}~\bibnamefont
  {Arkani-Hamed}}, \bibinfo {author} {\bibfnamefont {C.}~\bibnamefont
  {Figueiredo}}, \ and\ \bibinfo {author} {\bibfnamefont {F.}~\bibnamefont
  {Vaz{\~a}o}},\ }\href@noop {} {\  (\bibinfo {year} {2024}{\natexlab{d}})},\
  \Eprint {http://arxiv.org/abs/2412.19881} {arXiv:2412.19881 [hep-th]}
  \BibitemShut {NoStop}%
\bibitem [{\citenamefont {Arkani-Hamed}\ \emph
  {et~al.}(2018{\natexlab{b}})\citenamefont {Arkani-Hamed}, \citenamefont
  {Rodina},\ and\ \citenamefont {Trnka}}]{Arkani-Hamed:2016rak}%
  \BibitemOpen
  \bibfield  {author} {\bibinfo {author} {\bibfnamefont {N.}~\bibnamefont
  {Arkani-Hamed}}, \bibinfo {author} {\bibfnamefont {L.}~\bibnamefont
  {Rodina}}, \ and\ \bibinfo {author} {\bibfnamefont {J.}~\bibnamefont
  {Trnka}},\ }\href {\doibase 10.1103/PhysRevLett.120.231602} {\bibfield
  {journal} {\bibinfo  {journal} {Phys. Rev. Lett.}\ }\textbf {\bibinfo
  {volume} {120}},\ \bibinfo {pages} {231602} (\bibinfo {year}
  {2018}{\natexlab{b}})},\ \Eprint {http://arxiv.org/abs/1612.02797}
  {arXiv:1612.02797 [hep-th]} \BibitemShut {NoStop}%
\bibitem [{\citenamefont {Rodina}(2019{\natexlab{a}})}]{Rodina:2016mbk}%
  \BibitemOpen
  \bibfield  {author} {\bibinfo {author} {\bibfnamefont {L.}~\bibnamefont
  {Rodina}},\ }\href {\doibase 10.1007/JHEP09(2019)078} {\bibfield  {journal}
  {\bibinfo  {journal} {JHEP}\ }\textbf {\bibinfo {volume} {09}},\ \bibinfo
  {pages} {078} (\bibinfo {year} {2019}{\natexlab{a}})},\ \Eprint
  {http://arxiv.org/abs/1612.03885} {arXiv:1612.03885 [hep-th]} \BibitemShut
  {NoStop}%
\bibitem [{\citenamefont {Rodina}(2019{\natexlab{b}})}]{Rodina:2016jyz}%
  \BibitemOpen
  \bibfield  {author} {\bibinfo {author} {\bibfnamefont {L.}~\bibnamefont
  {Rodina}},\ }\href {\doibase 10.1007/JHEP09(2019)084} {\bibfield  {journal}
  {\bibinfo  {journal} {JHEP}\ }\textbf {\bibinfo {volume} {09}},\ \bibinfo
  {pages} {084} (\bibinfo {year} {2019}{\natexlab{b}})},\ \Eprint
  {http://arxiv.org/abs/1612.06342} {arXiv:1612.06342 [hep-th]} \BibitemShut
  {NoStop}%
\bibitem [{\citenamefont {Rodina}(2019{\natexlab{c}})}]{Rodina:2018pcb}%
  \BibitemOpen
  \bibfield  {author} {\bibinfo {author} {\bibfnamefont {L.}~\bibnamefont
  {Rodina}},\ }\href {\doibase 10.1103/PhysRevLett.122.071601} {\bibfield
  {journal} {\bibinfo  {journal} {Phys. Rev. Lett.}\ }\textbf {\bibinfo
  {volume} {122}},\ \bibinfo {pages} {071601} (\bibinfo {year}
  {2019}{\natexlab{c}})},\ \Eprint {http://arxiv.org/abs/1807.09738}
  {arXiv:1807.09738 [hep-th]} \BibitemShut {NoStop}%
\bibitem [{\citenamefont {Arkani-Hamed}\ and\ \citenamefont
  {Kaplan}(2008)}]{Arkani-Hamed:2008bsc}%
  \BibitemOpen
  \bibfield  {author} {\bibinfo {author} {\bibfnamefont {N.}~\bibnamefont
  {Arkani-Hamed}}\ and\ \bibinfo {author} {\bibfnamefont {J.}~\bibnamefont
  {Kaplan}},\ }\href {\doibase 10.1088/1126-6708/2008/04/076} {\bibfield
  {journal} {\bibinfo  {journal} {JHEP}\ }\textbf {\bibinfo {volume} {04}},\
  \bibinfo {pages} {076} (\bibinfo {year} {2008})},\ \Eprint
  {http://arxiv.org/abs/0801.2385} {arXiv:0801.2385 [hep-th]} \BibitemShut
  {NoStop}%
\bibitem [{\citenamefont {Britto}\ \emph
  {et~al.}(2005{\natexlab{a}})\citenamefont {Britto}, \citenamefont {Cachazo},\
  and\ \citenamefont {Feng}}]{Britto:2004ap}%
  \BibitemOpen
  \bibfield  {author} {\bibinfo {author} {\bibfnamefont {R.}~\bibnamefont
  {Britto}}, \bibinfo {author} {\bibfnamefont {F.}~\bibnamefont {Cachazo}}, \
  and\ \bibinfo {author} {\bibfnamefont {B.}~\bibnamefont {Feng}},\ }\href
  {\doibase 10.1016/j.nuclphysb.2005.02.030} {\bibfield  {journal} {\bibinfo
  {journal} {Nucl. Phys. B}\ }\textbf {\bibinfo {volume} {715}},\ \bibinfo
  {pages} {499} (\bibinfo {year} {2005}{\natexlab{a}})},\ \Eprint
  {http://arxiv.org/abs/hep-th/0412308} {arXiv:hep-th/0412308} \BibitemShut
  {NoStop}%
\bibitem [{\citenamefont {Britto}\ \emph
  {et~al.}(2005{\natexlab{b}})\citenamefont {Britto}, \citenamefont {Cachazo},
  \citenamefont {Feng},\ and\ \citenamefont {Witten}}]{Britto:2005fq}%
  \BibitemOpen
  \bibfield  {author} {\bibinfo {author} {\bibfnamefont {R.}~\bibnamefont
  {Britto}}, \bibinfo {author} {\bibfnamefont {F.}~\bibnamefont {Cachazo}},
  \bibinfo {author} {\bibfnamefont {B.}~\bibnamefont {Feng}}, \ and\ \bibinfo
  {author} {\bibfnamefont {E.}~\bibnamefont {Witten}},\ }\href {\doibase
  10.1103/PhysRevLett.94.181602} {\bibfield  {journal} {\bibinfo  {journal}
  {Phys. Rev. Lett.}\ }\textbf {\bibinfo {volume} {94}},\ \bibinfo {pages}
  {181602} (\bibinfo {year} {2005}{\natexlab{b}})},\ \Eprint
  {http://arxiv.org/abs/hep-th/0501052} {arXiv:hep-th/0501052} \BibitemShut
  {NoStop}%
\bibitem [{Note1()}]{Note1}%
  \BibitemOpen
  \bibinfo {note} {In all figures, we draw momentum polygons as disks for
  readability and convenience.}\BibitemShut {Stop}%
\bibitem [{Note2()}]{Note2}%
  \BibitemOpen
  \bibinfo {note} {For $n = 4$, only $c_{(1,5);(3,7)}$, $c_{(2,6);(4,8)}$,
  $c_{(2,8);(4,6)}$, and $c_{(2,4);(6,8)}$ do not satisfy the aforementioned
  criterion, and thus may not be determined by this simple trick. But, in the
  case at four-points, it is well-appreciated that gauge invariance is
  responsible for the contact term.}\BibitemShut {Stop}%
\bibitem [{Note3()}]{Note3}%
  \BibitemOpen
  \bibinfo {note} {One may hope that this procedure can determine \protect
  \textit {any} (non-leading-singularity) numerator as a function of numerators
  with one more pole. However, for this simple trick to work fully for the
  numerator of a term with $p$ poles, one requires a $(2p + 5)$-point
  amplitude. The same counting holds when we add $F^3$ and higher irrelevant
  terms to YM: in this case, equations analogous to \protect \cref
  {eq:c-det-2n} fully determine the contact term at $O(X^m)$ for amplitudes at
  $2m+1$ points and higher.}\BibitemShut {Stop}%
\end{thebibliography}%

\end{document}